\documentclass[aps,twocolumn,showpacs,preprintnumbers,nofootinbib,prd,superscriptaddress,groupedaddress,10pt]{revtex4-2}

\usepackage[utf8]{inputenc} 
\usepackage{graphicx,amssymb,amsmath,amsthm,amsfonts,epstopdf,epsfig,epsf,times}
\usepackage[linktocpage]{hyperref}
\usepackage[usenames]{color}
\usepackage{epstopdf}
\usepackage{textcomp}
\usepackage{bm}
\usepackage{latexsym}
\usepackage{rotating}
\usepackage{hyperref}
\usepackage{color}
\usepackage{longtable}
\usepackage{enumerate}
\usepackage{tensor}
\usepackage{stmaryrd}
\usepackage[normalem]{ulem}
\usepackage{mathtools}
\usepackage{url}
\usepackage{multirow}
\usepackage{graphicx}
\usepackage{mathtools}
\usepackage{verbatim}
\usepackage{soul,xcolor}
\setstcolor{red}

\definecolor{coolblack}{rgb}{0.0, 0.18, 0.39}
\definecolor{darkred}{rgb}{0.5,0,0}
\definecolor{darkgreen}{rgb}{0,0.5,0}
\definecolor{darkblue}{rgb}{0,0,0.5}
\definecolor{lapislazuli}{rgb}{0.15, 0.38, 0.61}
\definecolor{venetianred}{rgb}{0.78, 0.03, 0.08}
\definecolor{bleudefrance}{rgb}{0.19, 0.55, 0.91}
\definecolor{dogwoodrose}{rgb}{0.84, 0.09, 0.41}
\definecolor{dogwoodrose}{rgb}{0.84, 0.09, 0.41}
\definecolor{darkorgane}{rgb}{1,0.549,0}
\hypersetup{colorlinks=true, citecolor=darkblue, linkcolor=darkblue, 
urlcolor = darkblue}
\definecolor{olive}{rgb}{0.5, 0.5, 0.0}

\def\nn{\nonumber}

\newcommand{\ben}{\begin{enumerate}}
\newcommand{\een}{\end{enumerate}}

\def\be{\begin{equation}}
\def\ee{\end{equation}}
\newcommand{\beq}{\begin{eqnarray}}
\newcommand{\eeq}{\end{eqnarray}} 
\newcommand{\ba}{\begin{align}}
\newcommand{\ea}{\end{align}}

\def\be{\begin{equation}}
\def\ee{\end{equation}}
	
\newcommand{\bea}{\begin{eqnarray}}
\newcommand{\eea}{\end{eqnarray}}

\begin{document}
\title{Resonances, black hole mimickers and the greenhouse effect:\\ consequences for gravitational-wave physics}
%
%
\author{Vitor Cardoso$^{1,2}$, Francisco Duque$^{2}$
}
\affiliation{${^1}$ Niels Bohr International Academy, Niels Bohr Institute, Blegdamsvej 17, 2100 Copenhagen, Denmark}
\affiliation{${^2}$ CENTRA, Departamento de F\'{\i}sica, Instituto Superior T\'ecnico -- IST, Universidade de Lisboa -- UL,
Avenida Rovisco Pais 1, 1049 Lisboa, Portugal}

\begin{abstract} 
Ultracompact objects with photonspheres are known to mimic many observational features of black holes. It has been suggested that anomalous tidal heating or the presence of resonances in gravitational wave signals would be a clear imprint of a surface or absence of an horizon. Such claims and studies are all based on a frequency-domain analysis, assuming stationarity.
Here we show that the object needs to first ``fuel-up'' until it reaches the stationary regime.
The presence of a stable light ring and large light-travel times inside the object
may in fact delay enormously the ``charging-up'' and effectively contribute to the effacement of structure. In other words, black hole mimickers behave as black holes more efficiently than previously thought. Our results have implications for other resonant systems with sharp resonances, including ``floating orbits'' around spinning black holes. A proper accounting of the self-force for such systems seems to be both mandatory, and would have important applications for tests of horizon physics.
\end{abstract}
\maketitle


\section{Introduction} \label{sec:Intro}
The advent of gravitational-wave (GW) astronomy~\cite{LIGOScientific:2016aoc,Abbott:2020niy} and of very-long baseline interferometry~\cite{EventHorizonTelescope:2019dse,GRAVITY:2020gka} opened
exciting new windows to the invisible Universe~\cite{Barack:2018yly,Cardoso:2019rvt,Bertone:2018krk,Bar:2019pnz,Brito:2015oca}.
These observatories are tailored to study compact objects, such as black holes (BHs), and have therefore a tremendous discovery potential. The detection at large signal-to-noise ratios
of compact-binary inspirals allows for unprecedented tests of General Relativity in the strong-field regime~\cite{Barack:2018yly,Berti:2016lat,Cardoso:2019rvt,Bertone:2018krk,Brito:2015oca,Seoane:2021kkk}.

One of the foundational issues to address in the next decades concerns BH themselves: are they described well by General Relativity~\cite{Chrusciel:2012jk} in vacuum, and up to which extent are matter effects important and measurable~\cite{Cardoso:2016ryw,Barausse:2014tra,Cardoso:2021wlq}? At the core of these questions is the assumption that massive and dark objects are indeed BHs. Fortunately, GW astronomy and precise electromagnetic measurements allow us to probe the near-horizon region and to {\it quantify} the presence of horizons in the spacetime~\cite{Cardoso:2017cqb,Cardoso:2019rvt}.

It turns out that the dynamics -- at least at the perturbative level -- of very compact horizonless objects is only subtly different from that of BHs. When placed in a binary, a compact object will be tidally deformed by its companion, changing the multipolar structure of the gravitational field. Tidal love numbers of BHs vanish, whereas those of horizonless objects are small but nonzero~\cite{Cardoso:2017cfl,Maselli:2017cmm,LeTiec:2020bos}; the prompt ringdown of BHs, caused by trapping of GWs at the photonsphere, is common to all sufficiently compact objects, although late-time echoes should appear in horizonless geometries~\cite{Cardoso:2016rao,Cardoso:2016oxy,LIGOScientific:2020tif,Cardoso:2019rvt, Oshita:2018fqu, Oshita:2020dox, Oshita:2020abc}, and in fact these have already been searched in LIGO data with conflicting conclusions \cite{Abedi:2020sgg, Nielsen:2018lkf, Uchikata:2019frs, Abedi:2016hgu, Abedi:2018npz, LIGOScientific:2020tif,LIGOScientific:2021sio}.
Finally, compact and horizonless geometries absorb GWs in a substantially different way from BHs: the horizon is a one-way surface, whereas GWs are bound to escape from any horizonless alternative.
Thus, tidal heating can also be a good indicator of the BH nature of compact objects~\cite{Maselli:2017cmm,Datta:2019epe}.

It has been pointed out that horizonless compact objects would have low-frequency oscillation modes, which could be excited by orbiting bodies~\cite{Cardoso:2019nis,Maggio:2021uge,Fransen:2020prl,Fang:2021iyf,Sago:2021iku}.
This is in contrast with BHs, whose modes of oscillation are localized close to the photonsphere, where matter on stable orbits cannot exist~\cite{Berti:2009kk}. These resonances enhance GW emission, leading to a faster inspiral and a potentially-detectable dephasing with respect to BH spacetimes.

Previous analysis of detectability of resonance-crossing in BH mimickers were done in the frequency domain~\cite{Cardoso:2019nis,Maggio:2021uge,Fransen:2020prl,Fang:2021iyf,Sago:2021iku}.
By definition, the analysis assumes that the field is stationary, and then superposes an adiabatic evolution to evolve the particle in its motion, driven by GW emission.
The sole purpose of this work is to point out that the assumption of stationarity for BH mimickers may not and in fact does not hold in a large region of parameter space.
Very compact objects behave as a cavity, which have a very large ``build-up time,'' the time taken for arbitrary initial conditions to reach a stationary state. 
The build up time is of the order of the inverse of the resonance width itself, typically much larger than the timescale for evolution via GWs. Thus the conclusions of previous works need to be revisited. 

Since our analysis relies only on the existence of resonances, it is clear that our results have important implications for other systems. These include, for example, floating orbits around spinning BHs~\cite{Cardoso:2011xi,Zhang:2018kib}, or any other systems with sharp resonances. We note that our results and conclusions are consistent with classical results on electromagnetic cavities~\cite{Milton:2006ia}.

\section{Setup}\label{sec:Background}
\subsection{The theory}
We will work with a simple toy-model, that of a massless scalar field $\Phi$ around a compact horizonless object of mass $M$ in a spacetime background of metric $g_{\mu\nu}$, such that the line element is written as
\be
ds^2=-f dt^2+\frac{dr^2}{g}+r^2d\Omega_2^2\,,
\ee
where $d \Omega_2^2 = d\theta^2 + \sin^2 \theta d \varphi^2  $ is the metric on the two-sphere.

The scalar field will be excited by introducing  a point-like particle of mass $m_p$ coupled to it and orbiting around the central object. Letting $\tau$ denote the proper time of the point particle along the world line $z_p^\mu(\tau)=(t_p(\tau),r_p(\tau),\theta_p(\tau),\varphi_p(\tau))$, then its stress-energy tensor is given by 
\beq
T^{\mu\nu} = \frac{m_p}{\sqrt{-g}} \frac{dt}{d\tau}\frac{dz_p^\mu}{dt}\frac{dz_p^\nu}{dt}\frac{\delta(r-r_p(t))}{r^2}\delta^{(2)}(\Omega-\Omega_p(t)) \, , 
\eeq
and the full dynamics of the system is described by the action
\be
S[g,\Psi]=\int d^4x \sqrt{-g}\left(\frac{R[g]}{16\pi}-g^{\mu\nu}\partial_{\mu}\Psi\partial_{\nu}\Psi^*-2\alpha \Psi T\right)\,.
\ee
Here, $R[g]$ denotes the Ricci scalar of the metric, $\alpha>0$ is a coupling constant, and $T$ is the trace of the stress-energy tensor of the  particle.

We consider the point particle to act as a small perturbation. Thus, the background spacetime is fixed and taken to be the Schwarzschild exterior geometry and an internal geometry describing the compact horizonless object, with coordinates $\{t,r,\theta,\phi\}$. All that remains is to solve the scalar field equation of motion coupled to the point-like particle:
\beq
%
%
&&\frac{1}{\sqrt{-g}}\partial_{\mu}\left(\sqrt{-g}g^{\mu\nu}\partial_{\nu}\Psi\right)=\alpha T\,. \label{eq:KG}
\eeq
Throughout the rest of the paper we set $\alpha=1$ without loss of generality.


\subsection{The background}
For background, we use constant-density stars. Their interior is described by the metric functions~\cite{Shapiro:1983du}
\beq
f&=&\left(\frac{3}{2}\left(1-\frac{2M}{R}\right)^{1/2}-\frac{1}{2}\left(1-\frac{2Mr^2}{R^3}\right)^{1/2}\right)^2\,,\\
g&=&1-\frac{8\pi\rho}{3}r^2\,,
\eeq
where $R$ is the star's radius, $M$ is its mass, and $\rho=3M/(4\pi R^3)$ is its density.

The geometry above only describes ``realistic" stars when $R>9M/4$, otherwise the pressure diverges somewhere inside the star. Above some compactness, the geometry admits two light rings at the roots of $2f=rf'$~\cite{Cardoso:2008bp}. When $r<3M$, they are located at
\beq
r&=& 3M\,,\\
r&=&\frac{R\sqrt{4R^2-9MR}}{\sqrt{9MR-18M^2}}\,,
\eeq
where the first root coincides with the unstable light ring in vacuum Schwarzschild, and the second solution corresponds to a stable light ring located inside the star.
As discussed elsewhere, unstable light rings work as a trapping region slowly leaking energy out, and are responsible for the late time ringdown of gravitational wave signals involving BHs~\cite{Cardoso:2008bp,Cardoso:2019rvt}, or for the fading appearance of stars~\cite{1968ApJ...151..659A,Cardoso:2021sip}.
Stable light rings do not automatically imply linear instability, but they are linked to nonlinear instability mechanisms, which we won't discuss further~\cite{Keir:2014oka,Cardoso:2014sna}.
The (coordinate) frequency of light at the stable light ring as measured by a far-away observer is~\cite{Cardoso:2014sna}
\be
\Omega_{\rm LR2}=\frac{2\sqrt{M(R-9M/4)}}{R^2}\,,
\ee
corresponding to a transit time of $2\pi/\Omega_{\rm LR2}$ to a distant observer.

We can also calculate the transit time between the light ring and the center of the star, which dictates the period of trapped oscillations, and therefore of the ensuing echoes in the waveform~\cite{Cardoso:2016rao,Cardoso:2016oxy,Cardoso:2019rvt}. It turns out that this time, $T_{\rm light}$, is also approximately
$T_{\rm light}\approx 2\pi/\Omega_{\rm LR}$, although this may be a fortuitous aspect of very compact constant density stars \cite{Pani:2018flj}.

For definiteness, we will mostly focus on a configuration with $R=2.26 M$ and compare it with less compact geometries. This choice 
is close to the maximum possible compactness for this equation of state (the so-called Buchdahl limit), and the spacetime has two photonspheres,
being sufficiently compact to mimic some aspects of BHs.

\section{The build-up time of black hole mimickers\label{sec:buildup}}

\subsection{A scattering approach\label{subsec:greenhouse}}
%
\begin{figure}[t]
\begin{tabular}{c}
\includegraphics[width=0.9\columnwidth]{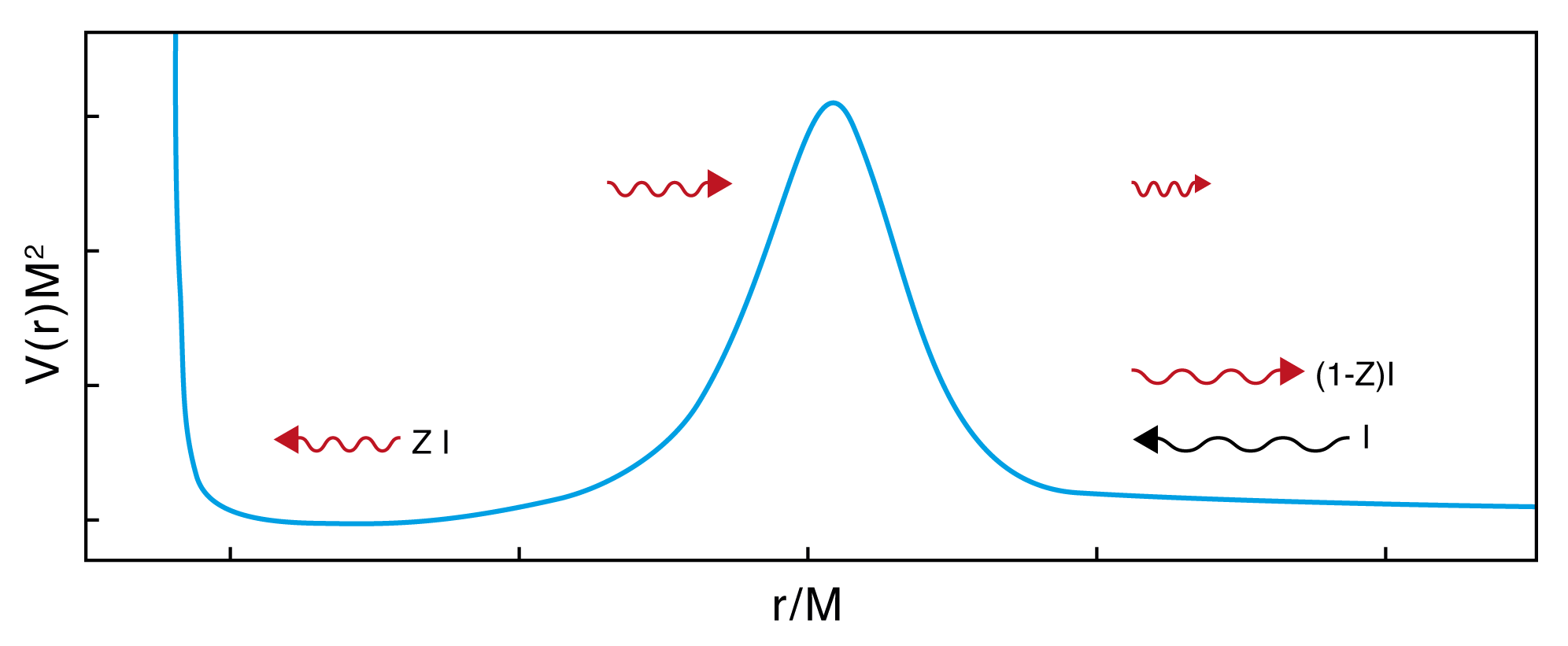}
\end{tabular}
\caption{Depiction of the effective potential governing massless fields on a horizonless but ultracompact geometry. The peak of the potential corresponds to the unstable photonsphere,
and the potential in its vicinities is indistinguishable from that of a BH spacetime. The centrifugal barrier in the object interior produces an effective cavity in the spacetime, from which waves slowly tunnel out or in. A cavity illuminated from the exterior ``heats-up'', in a manner akin to a greenhouse effect.}
\label{fig:potential} 
\end{figure}
An object which is sufficiently compact as to develop photonspheres is expected to behave as a cavity~\cite{Keir:2014oka,Cardoso:2014sna,Cardoso:2016rao,Cardoso:2016oxy,Cardoso:2017cqb,Cardoso:2019rvt}. Radiation is trapped in its interior, bouncing back and forth between its center or surface and the unstable photonsphere. The effective potential ``felt'' by massless fields is depicted in Fig.~\ref{fig:potential}. 

Consider now that such cavity is being bombarded by a constant flux of radiation ${\cal I}$ from a spin-$s$ wave carrying angular momentum $\ell$.
Such flux can correspond, for example, to that component radiated by an orbiting body which is directed towards the central object.
When it hits the barrier a small fraction $Z$ tunnels in, and one gets a reflected flux $(1-Z){\cal I}$. 
Here, the absorption coefficient $Z$ is frequency-dependent.
Denote by $T_0$ the roundtrip time of radiation inside the cavity.
Then after $T_0$, the radiation inside the barrier is now impinging from within, and a fraction $Z^2{\cal I}$ tunnels out, adding to the outgoing flux at large distances.
After a time $NT_0$ with $N$ an integer, one finds the outside flux to be
\be
{\cal F}_{NT_0}=(1-Z){\cal I}+Z^2 {\cal I}\sum_{j=0}^N(1-Z)^j ={\cal I}-{\cal I}Z(1-Z)^{N+1}\,.
\ee
This simple exercise predicts that the flux at large distances should be increasing in time steps of $T_0$ and relaxing to the final state on a timescale
\be
\tau=\frac{T_0}{Z}\,.\label{scattering_prediction}
\ee
The final state is that of an outgoing flux ${\cal I}$, as it should since the object is not absorbing.

The absorption factors can be worked out analytically in the low-frequency limit~\cite{Starobinski:1973,Starobinski2:1973,Brito:2015oca}
\begin{eqnarray}
Z_{slm} &=& C\left[\frac{(l-s)!(l+s)!}{(l!)^2}\right]^2 \,, \label{sigma}\\
C &=&4\left(2M\omega\right)^{2l+2}\left[\frac{(l!)^2}{(2l)!(2l+1)!!}\right]^2\prod_{k=1}^l\left[1+\frac{16M^2\omega^2}{k^2}\right]\,, \nn\\ \label{sigma0}
\end{eqnarray}
where $s$ refers to the spin of the perturbing field ($s=0$ for scalars, $s=-1$ for vectors and $s=-2$ for tensors) and $l,m$ refer to the multipolar mode of the perturbation.

Plugging in numbers, we find, for example,
\beq
Z_{010}&=&\frac{16M^4\omega^4}{9}\,,Z_{020}=\frac{64M^6\omega^6}{2025}\,,Z_{220}=\frac{256M^6\omega^6}{225}\,.\nonumber
\eeq
Thus, at low frequencies, the build-up time is very large and consequently any assumption on stationarity must be carefully justified, as we will see in practice below.
Notice the amusing fact that this calculation is similar to how the greenhouse effect for planet Earth is estimated in a naive approach.

Note finally, that the timescale \eqref{scattering_prediction} is the timescale that the system needs to ``settle,'' and is of the same order of magnitude as the resonant timescale, as implied by its quasinormal modes~\cite{Maggio:2018ivz,Cardoso:2019rvt}. In other words, a system composed of a compact object behaves as a cavity or resonator with lifetime~\eqref{scattering_prediction}.
\subsection{Resonances and forced oscillators\label{subsec:forcedoscillator}}
As discussed in the Introduction, we are interested in discussing possible resonances in our system, when the forced frequency $\omega$ induced by the orbital motion of the particle matches one of the natural frequencies $\omega_0$ of the compact object. To interpret our findings, it is useful to recall the results of a simple forced system with resonances, the forced harmonic oscillator (FHO). This is a particularly well suited description as we just saw that one can interpret the timescale to reach stationarity of a very compact object in terms of its resonant frequencies.

Take therefore a FHO described by the equation
\be
\frac{d^2\Psi}{dt^2}+\Gamma\frac{d\Psi}{dt}+\omega_0^2\Psi=F_0\cos\omega t\,,
\ee
with $F_0$ a force per unit mass.
The solution to the above which starts off at $\Psi (t=0)=\partial_t \Psi (t=0)=0$ is 
\beq
\Psi&=&F_0\frac{(\omega_0^2-\omega^2)}{(\omega_0^2-\omega^2)^2+\Gamma^2\omega^2}\left(\cos\omega t-e^{-\Gamma t/2}\cos\omega_\Gamma t\right)\nonumber\\
&+&F_0\frac{\Gamma \omega}{(\omega_0^2-\omega^2)^2+\Gamma^2\omega^2}\left(\sin\omega t-e^{-\Gamma t/2}\sin\omega_\Gamma t\right)\,,
\eeq
where $\omega_\Gamma=\sqrt{\omega_0^2-\Gamma^2/4}$.

From this simple analysis it is easy to see that when $\Gamma t\ll 1$ and for small damping $\Gamma\ll \omega_0$, the field grows on a timescale
\be
\tau_{\rm FHO}\approx \frac{2\pi}{\omega-\omega_0}\,.
\ee
This analysis is valid for short timescales and off the resonance. On resonance $\omega=\omega_0$, the field attains a maximum on a timescale of $\tau\sim 1/\Gamma$.
As we suggested above, for compact horizonless objects one should identify $1/\Gamma$ ($\Gamma$ is intrinsic to the resonating system and corresponds roughly to its mode $\omega_I$ as we detail below, see Eq.~\eqref{EQ_qnm_def}) with the timescale \eqref{scattering_prediction}.
\section{Numerical Results}
We now wish to show, via explicit numerical examples, that the previous picture is correct, and that the handling of resonances must be made carefully,
specially when done in the frequency domain.
\subsection{A point particle orbiting a compact object}
%
\begin{table}[ht!]
\begin{tabular}{c c c} \hline\hline
\multirow{2}{*}{$M\omega_\text{QNM}$}
& \multicolumn{2}{c}{$r_p/M$}  \\ 
\cline{2-3}
& $a=0M$ & $a=0.9M$ \\
\hline \hline
$0.0881 - i1.197\times 10^{-7}$& 5.051 & 4.780 \\
$0.1259 - i2.687\times 10^{-6}$& 3.981 & 3.674 \\
$0.1633 - i2.470\times 10^{-5}$& 3.347 & 3.011\\
\hline\hline
\end{tabular}
\caption{The lowest $l=1$ scalar quasinormal frequencies of a uniform-density relativistic star with $R=2.26M$. We also show the corresponding orbital radius at which the mode would be excited, calculated by equating the orbital frequency $\Omega$ in Eq.~\eqref{eq:AngularFreq} to the real part of the QNM frequency and solving for $r_p$.
The value of $a$ corresponds to that (ficticious spin) used in expression for the orbital frequency $\Omega$ (cf. Eq.~\eqref{eq:AngularFreq}). For less compact stars, resonant frequencies are impossible to excite with matter on circular orbits. For example, for $R=6M$ the lowest dipolar QNM frequency is $M\omega=0.262189 - i0.204880$. 
}
\label{tab:QNMs}
\end{table}
We start by placing a pointlike particle in circular orbit around a constant-density star,
\beq
r_p=\text{const} \, ,\quad \theta_p =\frac{\pi}{2} \, ,\quad \varphi_p = \Omega\, t \,.
\eeq
For non-rotating compact objects, it is well-known that the Schwarzschild geometry admits stable timelike circular geodesics for radius larger than the innermost circular stable orbit (ISCO) at $r_\text{ISCO}=6M$~\cite{chandrasekhar1992mathematical}. Their orbital frequency is $\Omega=\sqrt{M/r_p^3}$. While they can excite some proper modes of very compact constant-density stars (reference values are shown in Table~\ref{tab:QNMs}), the timescales of these resonances are too large to be probed by our numerical setup in a reasonable time frame with sufficient accuracy. The only possibility would be to consider unstable circular geodesics, which have larger frequencies and are able to excite modes which grow on smaller timescales. However, since we eventually want to understand the impact of energy loss on the orbit, unstable motion is not an option.

The only remaining possibility is to consider non-geodesic motion. Thus, we consider a particle following non-geodesic motion around a non-rotating compact object. To make the motion as simple as possible and still satisfy the requirement that it excites resonant modes, we consider the motion to be equivalent to that around a Kerr BH with mass $M$ and spin $a$. In this case, the angular frequency $\Omega$, energy $E$ and angular momentum $L_z$ of the orbital motion are
\beq
\Omega &=& \frac{\sqrt{M}}{r_p^{3/2} + a\sqrt{M}} \, , \label{eq:AngularFreq} \\
\epsilon&\equiv&\frac{E}{m_p}=\frac{r_p^{3/2} - 2M r_p^{1/2}+ a \sqrt{M}}{r_p^{3/4}\sqrt{r_p^{3/2} - 3M r_p^{1/2}   + 2 a \sqrt{M} }  } \, , \label{eq:Energy}\\
\mathcal{L}_z&\equiv&\frac{L_z}{m_p} = \frac{\sqrt{M}\left(r_p^2 - 2 a \sqrt{M}\,r_p^{1/2} + a^2 \right)}{r_p^{3/4}\sqrt{r_p^{3/2} - 3M r_p^{1/2}   + 2 a \sqrt{M} }  } \label{eq:Lz},
\eeq
where $0\leq a/M \leq 1$ is now a free ``knob'' (which, were the central object a spinning BH, would be the BH spin). We note that, while this is not geodesic motion,
the sole purpose of prescribing such motion is to allow us to investigate resonances and resonance-crossing in realistic timescales with acceptable accuracy. 
The nature of the motion is not relevant to the resonances we want to focus on.

As we show in Appendix~\ref{sec:Unstable}, the timescale associated with the excitation of the resonance is independent of this choice, being completely controlled by the frequency of the circular orbit. Our imposition of this artificial motion is purely pragmatic, as this is a simple way to make stable circular orbits have higher frequency without having to change the geometry of the central object. \footnote{Non-geodesic motion may (and will, in this setup) excite also the modes of a non-rotating black hole, contradicting our claims in the introduction. Note however, that this is a pure artifact of non-geodesic motion, which was also seen in triple systems~\cite{Cardoso:2021vjq}. However, these modes are much short-lived in comparison with the low-frequency modes of horizonless geometries that we are interested in; hence we will not investigate this issue further.}

To calculate the field amplitude and energy fluxes, we used two different numerical schemes. One works in the time domain, and it smoothens the pointlike character of the orbiting object~\cite{Krivan_1997,LopezAleman:2003ik,Pazos_valos_2005,Sundararajan:2007jg,Cardoso:2019nis,Cardoso:2021sip}. 
The other technique is based on separation of angular variables using spheroidal harmonics~\cite{Berti:2005gp} in the frequency domain, where one can apply standard Green function techniques~\cite{Davis:1971gg,Mino:1997bx,Cardoso:2002ay,Berti:2010ce}. The latter assumes a stationary profile and is used to benchmark our results at asymptotically late times. Both approaches are well documented and have been widely tested in the past. We also note that similar techniques have been employed in the past for the problem of particle scattering by the same relativistic constant density stars introduced in Section~\ref{sec:Background} \cite{Kokkotas:1995av,Andrade:1999mj,Ferrari:2000sr,Tominaga:2000cs}, where a transient excitation of quasinormal modes can also be observed.

For the time-domain code, we use as initial data
\be
\Psi |_{t=0} =\partial_t \Psi |_{t=0} = 0 \,.
\ee
Such initial data leads in general to an initial burst of energy which has no implication in the long term results we discuss here.
Also, for both codes we compute the scalar energy flux $\mathcal{F}$ emitted to far-away distances through
\beq
\mathcal{F}=\frac{dE}{dt}=\lim_{r\to\infty}\frac{1}{r^2}\int d\Omega\, \partial_t \Psi\, \partial_r \Psi \, .
\eeq
This quantity scales with $m_p^2$ and we will often choose to work in terms of the scale invariant energy flux $q^{-2} \mathcal{F}$, where $q=m_p/M$ is the mass ratio of the binary.

\begin{figure*}[ht!]
\begin{tabular}{c}
\includegraphics[width=0.66\columnwidth]{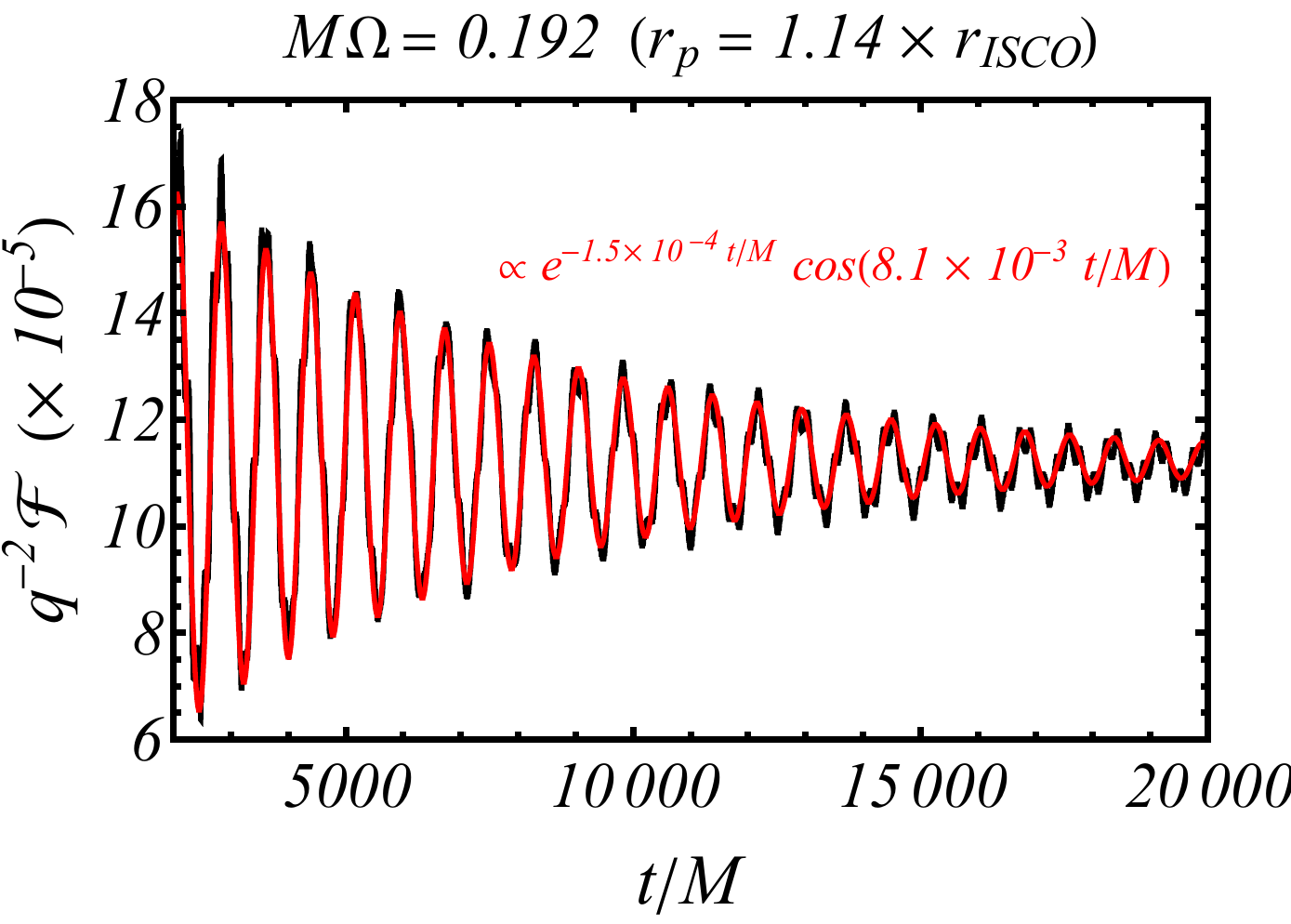}
\includegraphics[width=0.66\columnwidth]{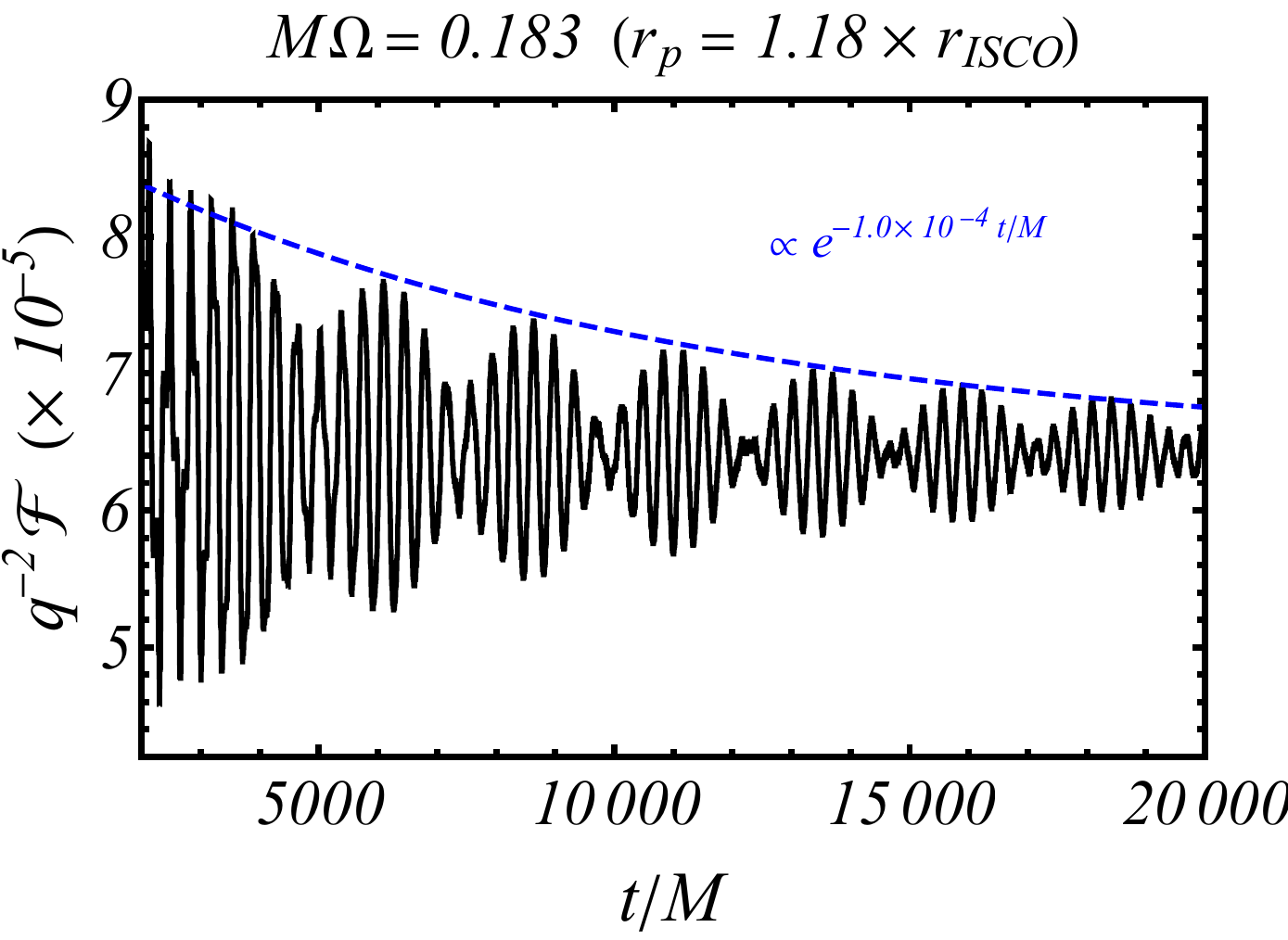}
\includegraphics[width=0.66\columnwidth]{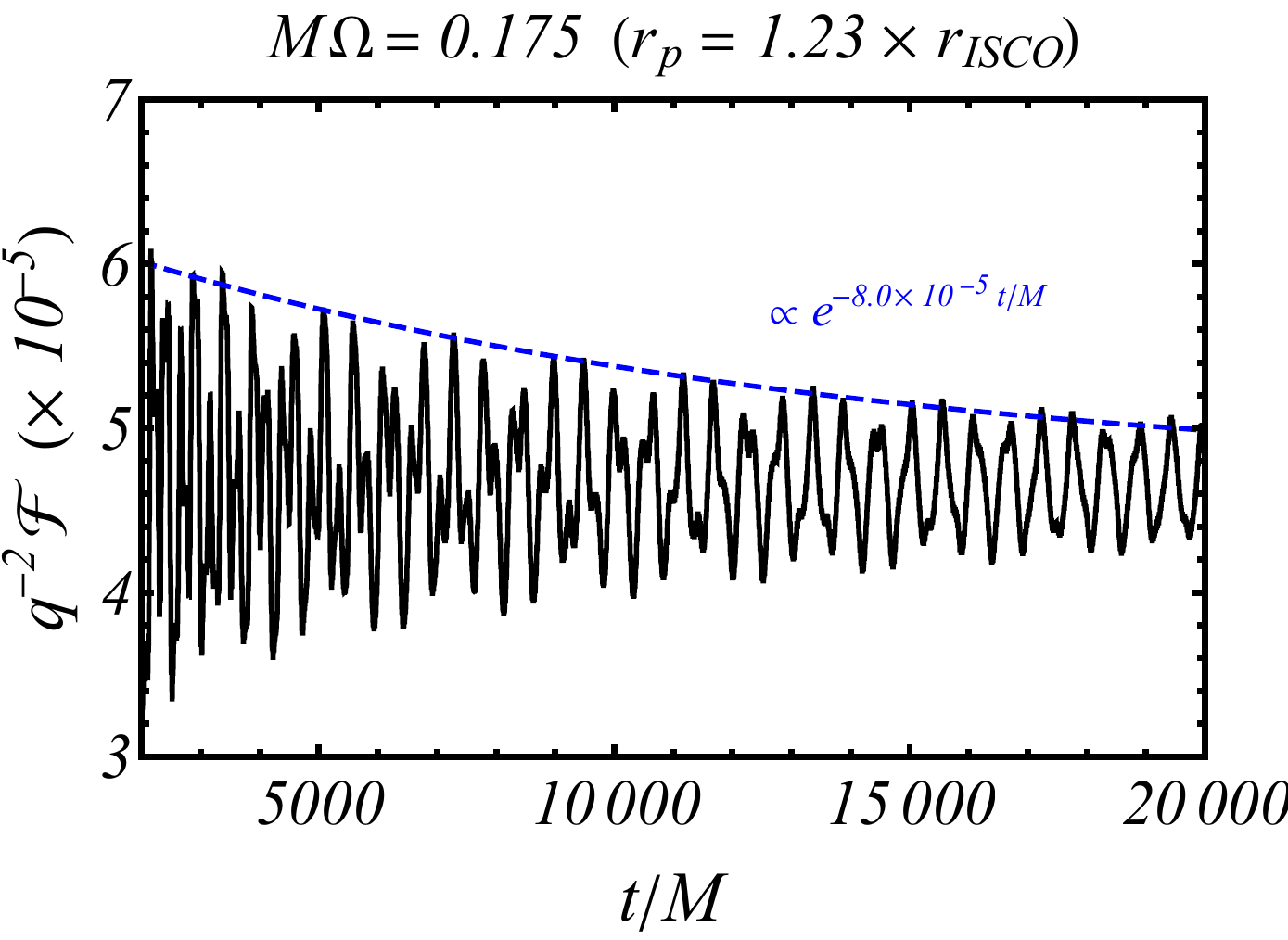} \\
\includegraphics[width=0.66\columnwidth]{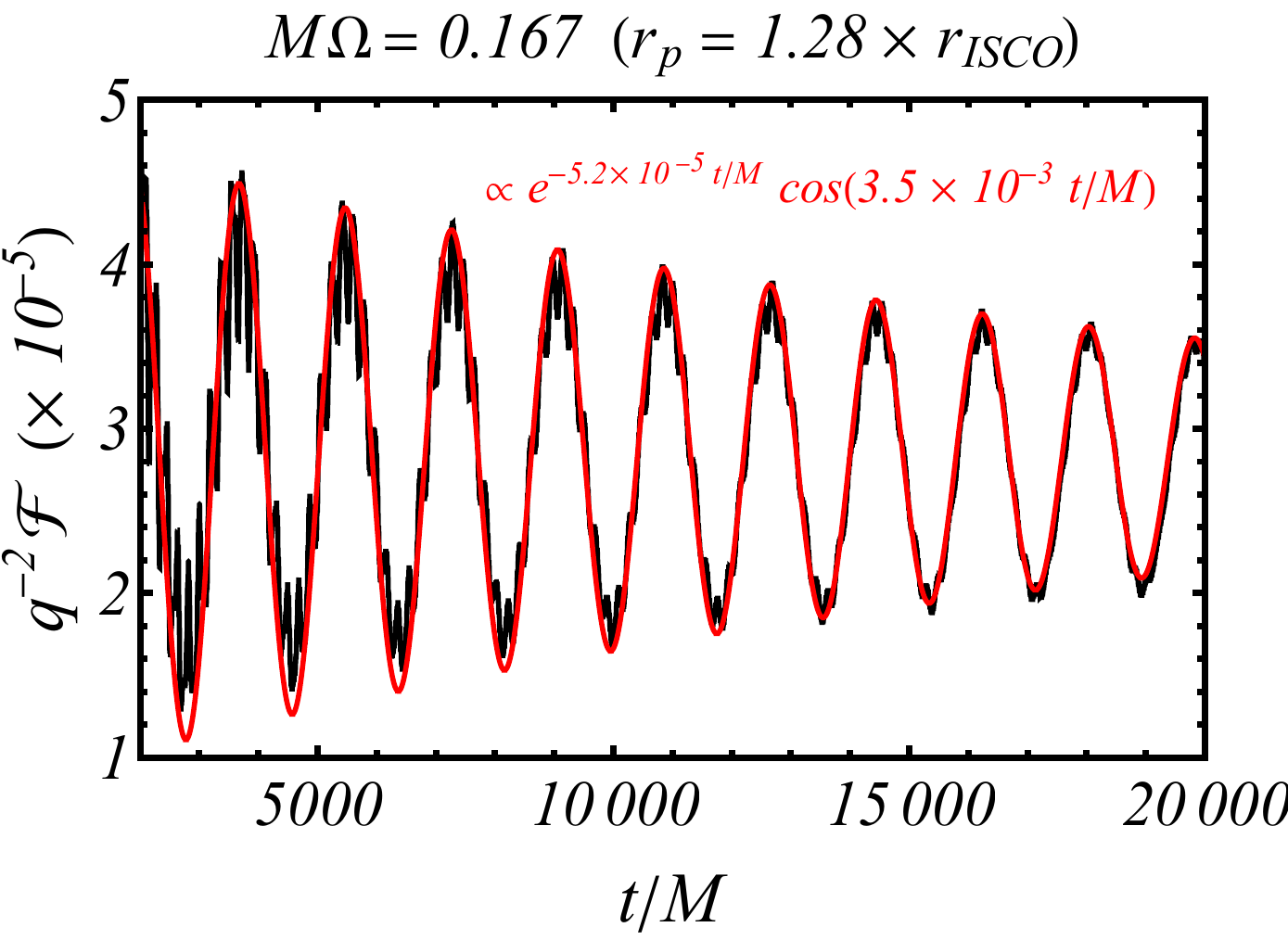} 
\includegraphics[width=0.66\columnwidth]{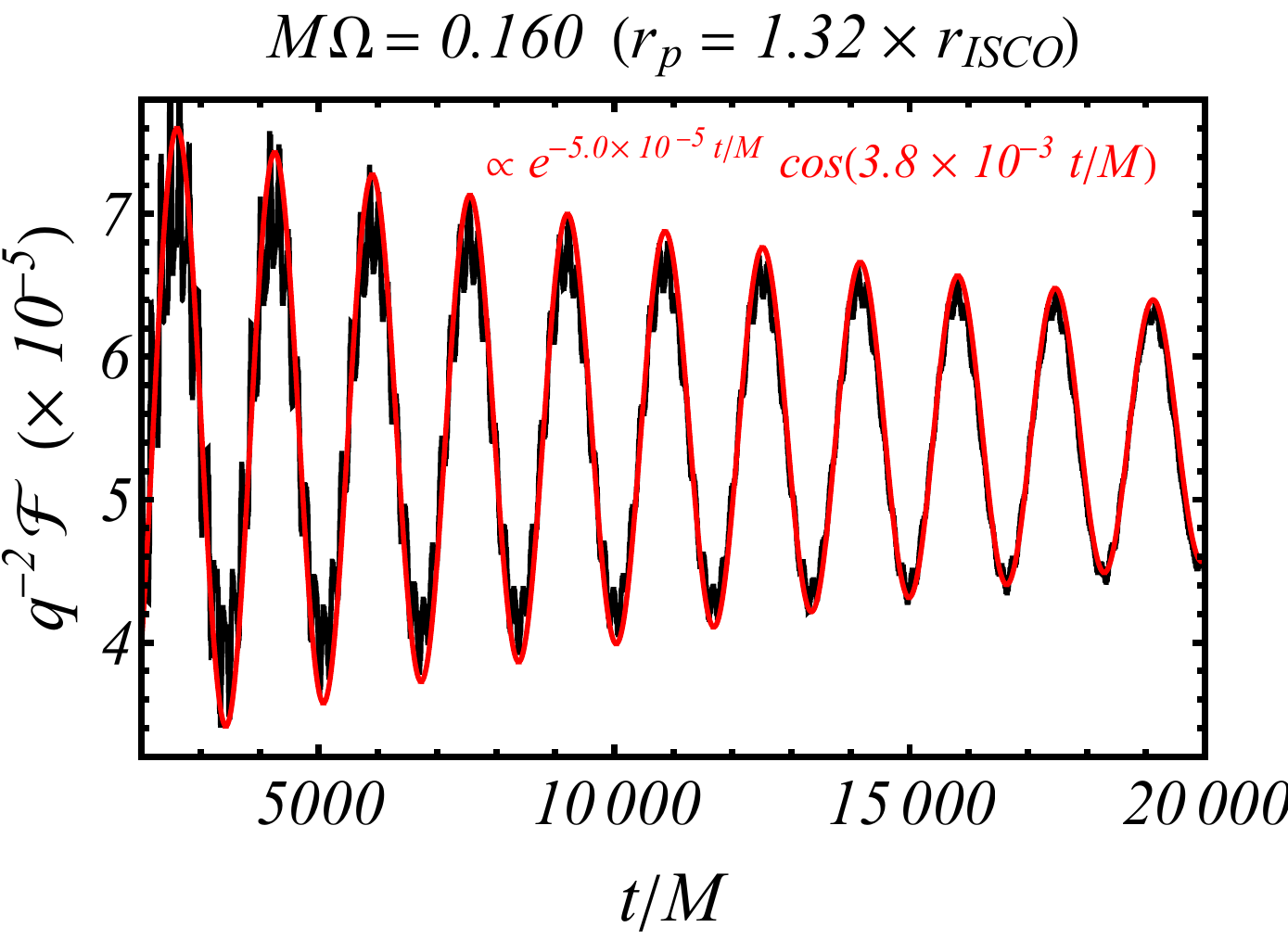} 
\includegraphics[width=0.66\columnwidth]{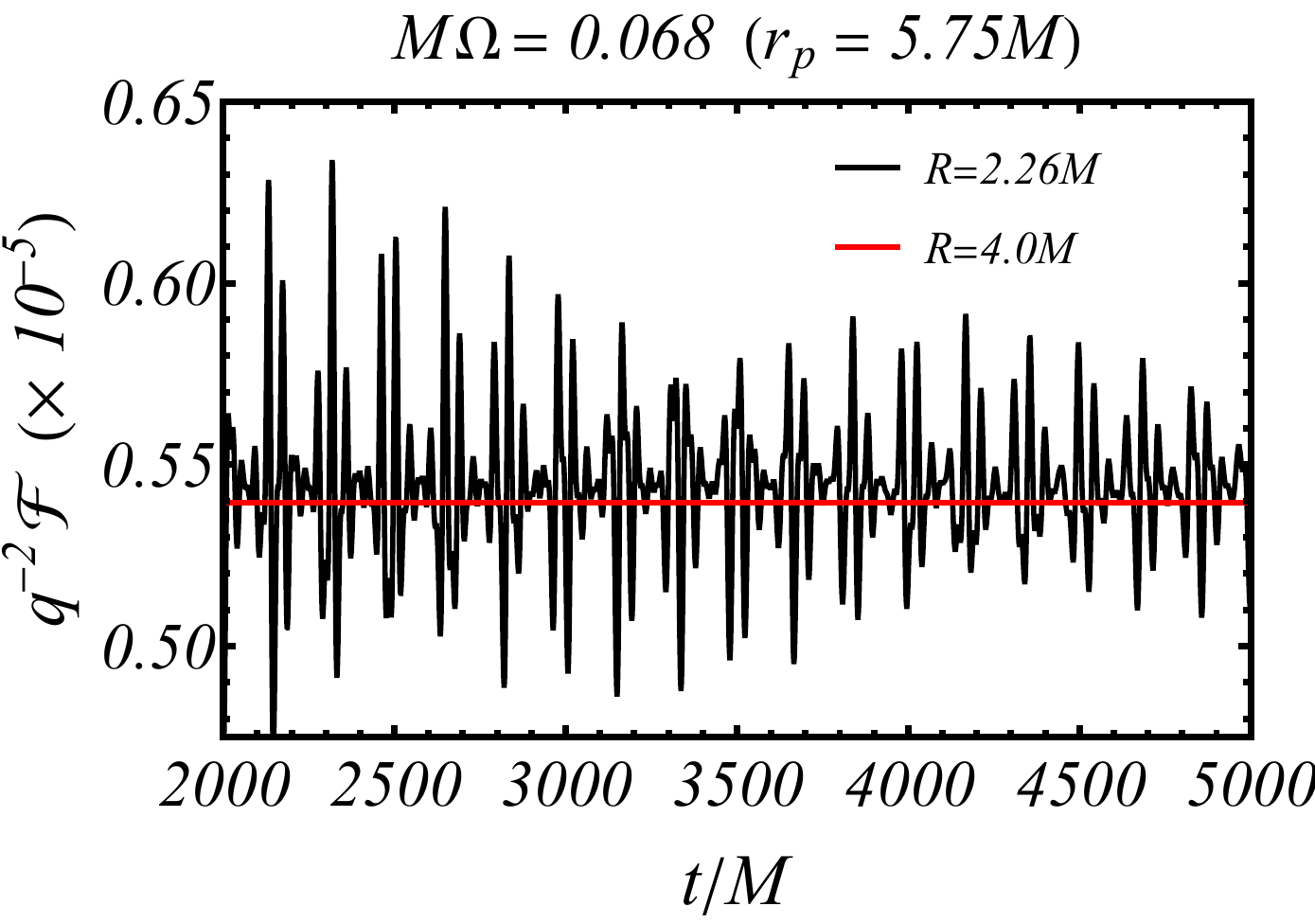} 
\end{tabular}
\caption{Evolution of the scalar energy flux $\mathcal{F}$ by a point-particle of mass $m_p$, made to orbit a constant-density star of mass $M$ on a circular orbit of constant radius $r_p$ (the orbit is not allowed to evolve). We normalize the flux by the mass ratio $q=m_p/M$. The results refer to the dipolar mode ($l=1$), but results are similar for higher multipoles. 
Except for the right bottom panel, the star has radius $R=2.26M$, and the frequency $\omega$ corresponds to the angular frequency $\Omega$ of the circular orbit, with $a=0.9M$ in Eq.~\eqref{eq:AngularFreq} ($r_\text{ISCO}\approx2.321M$). At late times, the flux asymptotes to a constant which agrees with the value computed in the frequency-domain. The relaxation time is large for stars with photonspheres, but very short for less compact stars, where the system quickly becomes stationary, as seen in the right-bottom panel. 
}
\label{fig:l1} 
\end{figure*}
%
\subsection{The build-up time}\label{sec:buildup_numerical}

\begin{figure}[ht!]
\begin{tabular}{c}
\includegraphics[width=0.9\columnwidth]{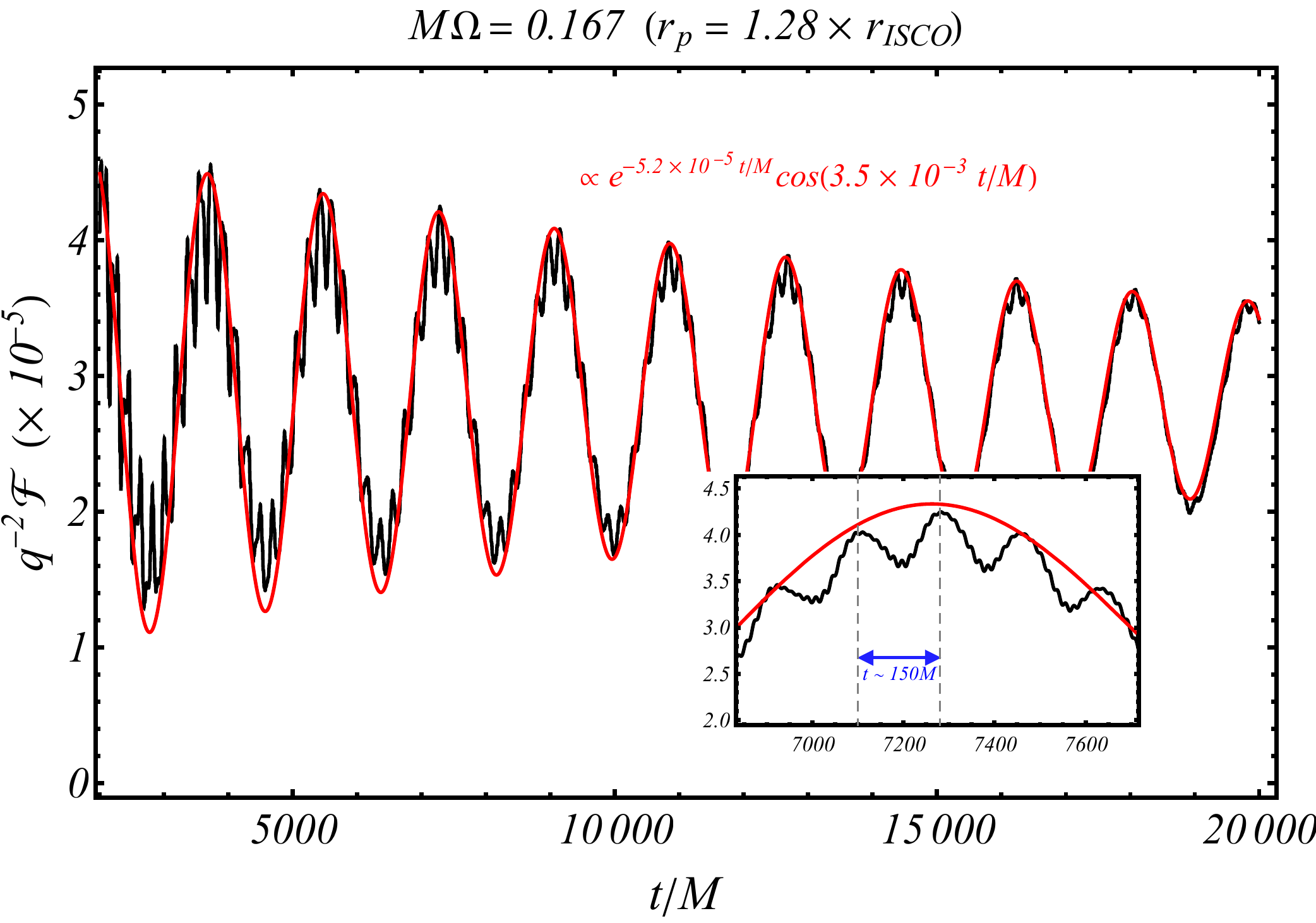} 
\end{tabular}
\caption{Scalar energy flux by a point particle in circular orbit at $r_p=1.28\, r_\text{ISCO}$, with angular frequency $\Omega$ given by Eq.~\eqref{eq:AngularFreq} with $a=0.9M$, around a constant density star of radius $R=2.26M$ (the orbit is not evolving, the particle remains at fixed $r_p$). There are three different timescales in the signal: a high-frequency component corresponding to the ``direct signal'' with an orbital period $T/2= \pi / \Omega \sim 19M $ (the $1/2$ factor appears since we are showing fluxes); the traveling time $~150M$ of waves inside the cavity potential; a lower frequency ``envelope'' corresponding to the excitation of the QNM of the constant density star with frequency $M\omega_\text{QNM}=0.16333 - i 2.470 \times 10^{-5}$. This leads to a beating whose frequency is given by the semi-difference between the orbital and the QNM frequency $2\pi/(\Omega - \omega_\text{QNM}) \sim 1800M $.}
\label{fig:Inset} 
\end{figure}

We take $a=0.9M$ in Eq.~\eqref{eq:AngularFreq} because it is one of the smallest values for the spin that allow us to probe a fast-growing resonance, while keeping the circular motion stable.
Our time-domain numerical results are summarized in Figs.~\ref{fig:l1}-\ref{fig:Inset}. 

In spacetimes without trapping regions, in this context without photonspheres, the initial data relaxes on a few dynamical timescales to a final stationary result, which coincides with that obtained via a frequency-domain approach. This behavior is apparent for the dipolar mode of an $R=4M$ uniform-density star in Fig.~\ref{fig:l1} (results are similar for the quadrupolar mode).

By contrast, for spacetimes which are sufficiently compact as to have photonspheres, the approach to stationarity is a long process. As explained above in Section~\ref{sec:buildup},
the photonsphere is responsible for a potential barrier, through which waves need to tunnel and ``build-up'' until a stationary state is reached. The very first stages of this process are -- in accordance with the analysis of Section~\ref{subsec:greenhouse}-- a slow growth of the outgoing flux in steps of $T_0$, the light travel time inside the photonsphere (see also the inset of Fig.~\ref{fig:Inset}, where such steps are clear). A few other features are worth highlighting. The relaxation time is in good agreement with the analysis of Section~\ref{subsec:greenhouse}.
For $R=2.26M$ and $r_p=1.14 \, r_\text{ISCO}$ -- $M\omega_{l=1} = 0.193$  as set by Eq.~\eqref{eq:AngularFreq} -- our numerical results indicate a relaxation timescale $\tau \sim 6500M, 4500M$ for $l=1,\,2$, whereas Eq.~\eqref{scattering_prediction} would indicate $\tau \sim 4000M, 3100M$ respectively (we evaluated the amplification factors numerically). The relaxation time increases when the circular orbit radius increases, again in line with prediction~\eqref{scattering_prediction}.
The proper oscillation modes of cavities govern to a good extent their dynamical behavior, so we calculated the QNMs of scalar fields in a constant-density star with known methods~\cite{Berti:2009kk,GRITJHU}. They are shown in Table~\ref{tab:QNMs}. The QNM frequencies are located in the complex plane, and we write them as
\be
\omega_{\rm QNM}=\omega_R+i\omega_I\,.\label{EQ_qnm_def}
\ee

Our results also show finer details, in particular beatings and finer structure at small timescales, apparent in Fig.~\ref{fig:l1}. A zoom-in for $M\omega=0.167M$ ($r_p=1.28r_\text{ISCO}$) is shown in Fig.~\ref{fig:Inset} for the dipolar mode. These features can be understood with the three different scales of the problem: the orbital timescale, $T/2= \pi / \Omega \sim 19M$ shows up as the smallest timescale in the problem and is clear in the inset of Fig.~\ref{fig:Inset} (the $1/2$ factor appears since we are discussing fluxes); the orbital frequency $M\Omega=0.1668$ is close to the resonant QNM frequency $M\omega_{R}=0.1633$ (see Table~\ref{tab:QNMs}). The forced harmonic oscillator of Section~\ref{subsec:forcedoscillator} then anticipates a beating mode of frequency $\Omega-\omega_{R}$, i.e., a beating period $\tau_{\rm beating}\sim 1800M$, in good agreement with our numerics. Notice also that steps of $\sim 150M$ are clear in Fig.~\ref{fig:Inset}, which correspond to the travel time of waves inside the cavity, hence to the buildup of the field in the cavity.

\begin{figure}[ht!]
\begin{tabular}{c}
\includegraphics[width=0.9\columnwidth]{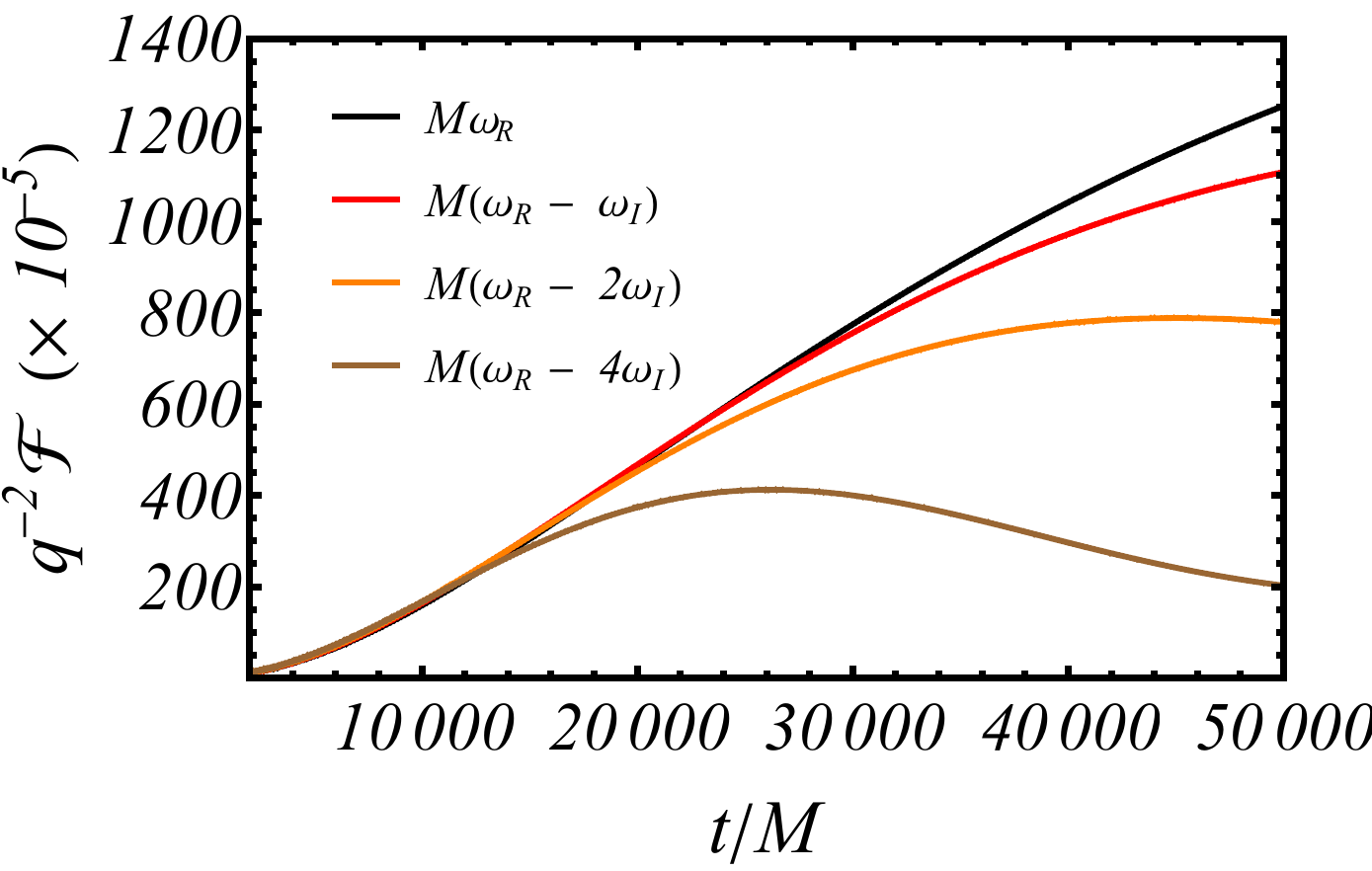} 
\end{tabular}
\caption{Resonant excitation of the dipolar QNM of a constant-density star of radius $R=2.26M$ with frequency $M\Omega=M\omega_R=0.1633$ (cf. Table~\ref{tab:QNMs}), corresponding to a point particle at $r_p=3.011M$. A small deviation of this radius resulting in frequency shift of $\delta \Omega \gtrsim 2 \omega_I$ can significantly hinder the excitation of the resonance. This is in agreement with standard results for the driven-harmonic oscillator, where the frequency bandwith of the resonance peak is $\delta \Omega\sim \omega_I$.}. 
\label{fig:Resonance} 
\end{figure}
Finally, to excite the resonance, one needs to tune to frequencies closer to the resonant QNM. Our results are shown in Fig.~\ref{fig:Resonance}, again for the dipolar mode.
The flux reaches amplitudes which are two orders of magnitude larger, but large timescales ${\rm min}\left(1/\omega_I, 2\pi/(\Omega-\omega_R)\right)$ are required for this build-up. The frequency needs to be very fine-tuned in order to properly excite the resonance, since as expected from the driven-harmonic oscillator, the bandwidth of the resonance peak is $\delta \omega \sim \omega_I$. Hence, when $\omega_I$ is very small as it happens for the proper modes of horizonless ultracompact objects, the region of the parameter space where the resonance can be triggered is limited and in addition the resonance takes a lot of time to develop. As we will see below, the combination of the two conditions can jeopardize the ability to effectively excite a resonance in a binary system.

One could question the generality of our results considering the artificial motion that we took for the point particle. However, in Appendix~\ref{sec:Unstable} we repeat the analysis for $a=0M$, which makes motion geodesic. By placing the particle at radii that yields the same orbital angular frequency as the ones presented in Fig.~\ref{fig:l1}, we observe that the timescales involved are exactly the same for every single case, with only the relative magnitude between the fluxes changing. Note that in order to excite the quasinormal mode with $M\omega_\text{QNM}=0.16333 - i2.470\times 10^{-5}$ with a circular geodesic, the point particle would have to be placed at $r_p=3.347M$, i.e., at an unstable orbit. This would impede the study of inspiralling trajectories if we were using this setup.

\section{Inspiralling objects and consequences for gravitational-wave physics}
What we have shown is that very compact objects are good ``absorbers'' of GWs for a limited but long amount of time. 
This is a pure spacetime geometry effect, whereby GWs get trapped within the photonsphere. The trapping takes a timescale of the order of \eqref{scattering_prediction},
after which radiation is re-emitted. These objects then recycle GWs with a certain delay time.
Thus, the physics of these objects must take into account such delay.

What we argue now is that there are two main effects, none of which was considered with due care in the literature.
The first concerns the dynamics away from the resonances. When these dynamics happen on short timescales -- such as the final stages of the coalescence -- then
the cavity has not time to ``fuel up'' and absorbs most of the impinging radiation: horizonless compact objects then behave to a good approximation as BHs, with equivalent absorption
properties, and possibly indistinguishable from them. 
The second effect concerns the crossing of resonances, a rather generic effect not particular to compact horizonless objects: we show that frequency-domain evolutions do not
capture the entire physics and must be complemented by addition constraints, when time evolutions are prohibitive.
\subsection{Adiabatic evolution of orbits and energy balance}

To study GW-driven inspirals, we consider adiabatic evolutions, where the point particle is always on a circular orbit with some associated energy and angular momentum. We place the particle at some initial radius $r_0$, and determine the initial energy and angular momentum according to Eqs.~\eqref{eq:Energy}-\eqref{eq:Lz}. Then, we need to evaluate the backreaction on these, due to emission of energy (and angular momentum). As we argue in this work, the flux needs to include the energy loss to infinity, but it should also include the energy piling up within the cavity.
However, considering effects of the cavity is a challenging problem which we won't address here. In this work, we will only take into account the energy radiated away to infinity, but we insist that the cavity may play an important role. For circular orbits, the angular momentum net balance is completely determined by the energy balance so we only need to solve
\beq
\frac{dE}{dt}=-\mathcal{F} \, , 
\eeq
with the appropriate initial energy, and use this to evolve
\beq
\frac{dr}{dt}=-\mathcal{F} \left(\frac{dr}{dE} \right) \, , 
\eeq
again with the appropriate initial conditions. Having the updated value for $r_p$, we can compute the angular frequency $\Omega$ again using
\beq
\frac{d\Omega}{dt}=\left(\frac{dr}{dt}\right) \left(\frac{d\Omega}{dr} \right) \, .
\eeq

This procedure can be applied both for the time and the frequency-domain. However, the flux computed in the frequency-domain implicitly assumes stationarity, i.e. that the oscillations around the average flux vary out to zero much faster than the timescale on which the particle inspirals. For the systems we are discussing, this implies the cavity has had time to fuel-up. For the time-domain instead, the energy balance is done at every instant and therefore can account for the inhomogeneities in the flux as the star is relaxing or the cavity is fueling-up.  

\subsection{Off resonance}
The results above have important implications for GW emission. As the small object of mass $m_p$ inspirals, the frequency is changing. For objects on a quasi-circular orbit millions of years prior to merger, a ``stationary state'' (to be read as where the frequency-domain calculation yields the same result as time-domain) is reached. However, in the late stages of inspiral, the frequency is varying rapidly and hence not allowing the compact object to ``fuel up''. This happens whenever the frequency change
\be
\Delta \omega \gtrsim \omega_R\,,
\ee
and the corresponding inspiral time is small enough that doesn't allow for relaxation, $t_{\rm inspiral}<\tau$ (cf. Eq.~\eqref{scattering_prediction}). Let's start the process at some $r_p(t=0)=r_0$. Then, for quasi-circular orbits, and including only GW reaction~\cite{Peters:1964zz}
\beq
r_p(t)&=& (r_0-4\beta t)^{1/4}\,,\\
\beta&=&\frac{64}{5}M^2m_p\,.
\eeq
The time taken to inspiral from $r_0$ to $r_p(t)$ can also be written in terms of the initial and final GW frequency $\omega_0,\omega_f$ as
\be
t_{\rm inspiral}=\frac{2^{2/3}M^{4/3}\left(1-(\omega_0/\omega_f)^{8/3}\right)}{\beta \omega_0^{8/3}}\,.
\ee

We thus find, 
\be
\frac{t_{\rm inspiral}}{\tau}\sim 10^{-2} \frac{100M}{T_0}\frac{10^{-5}M}{m_p}\left(\frac{M\omega}{0.06}\right)^{10/3}\,.
\ee
In other words, cavity effects during extreme mass ratio inspirals should be taken into account. Time or frequency domain analysis
should then include the temporary pile up of energy in the cavity as the inspiral progresses. We will not dwell on this important topic here, except to highlight two aspects.
The first is that a proper analysis should take into account radiation reaction, by a proper modeling of the entire system. It's a challenging problem, but one that would bear many fruits.
The second aspect is that the above also shows how the BH limit is approached in a natural way. The BH limit can be thought of as the $T_0\to \infty$ limit of the previous construction.
In this limit, the central object is a perfect absorber during the entire inspiral, and the delay time between receiving and returning radiation diverges. Therefore, a proper handling of the cavity problem should recover the BH result continuously.

Note that previous works suggested that the cavity would only be important for the evolution of the binary when the travelling time inside it is comparable (or larger) than the radiation-reaction timescale \cite{Maselli:2017cmm, Maggio:2021uge}. However, as discussed above, energy can be trapped by being reflected back and forth in the cavity until it saturates. This process corresponds to multiple travel times, as dicated by Eq.~\eqref{scattering_prediction}. In general, this timescale can be much bigger than the travel time inside the cavity, which makes the latter more relevant for larger mass-ratios than what was previously considered.
\subsection{Crossing resonances}
%
\begin{figure}[t]
\begin{tabular}{c}
\includegraphics[width=0.95\columnwidth]{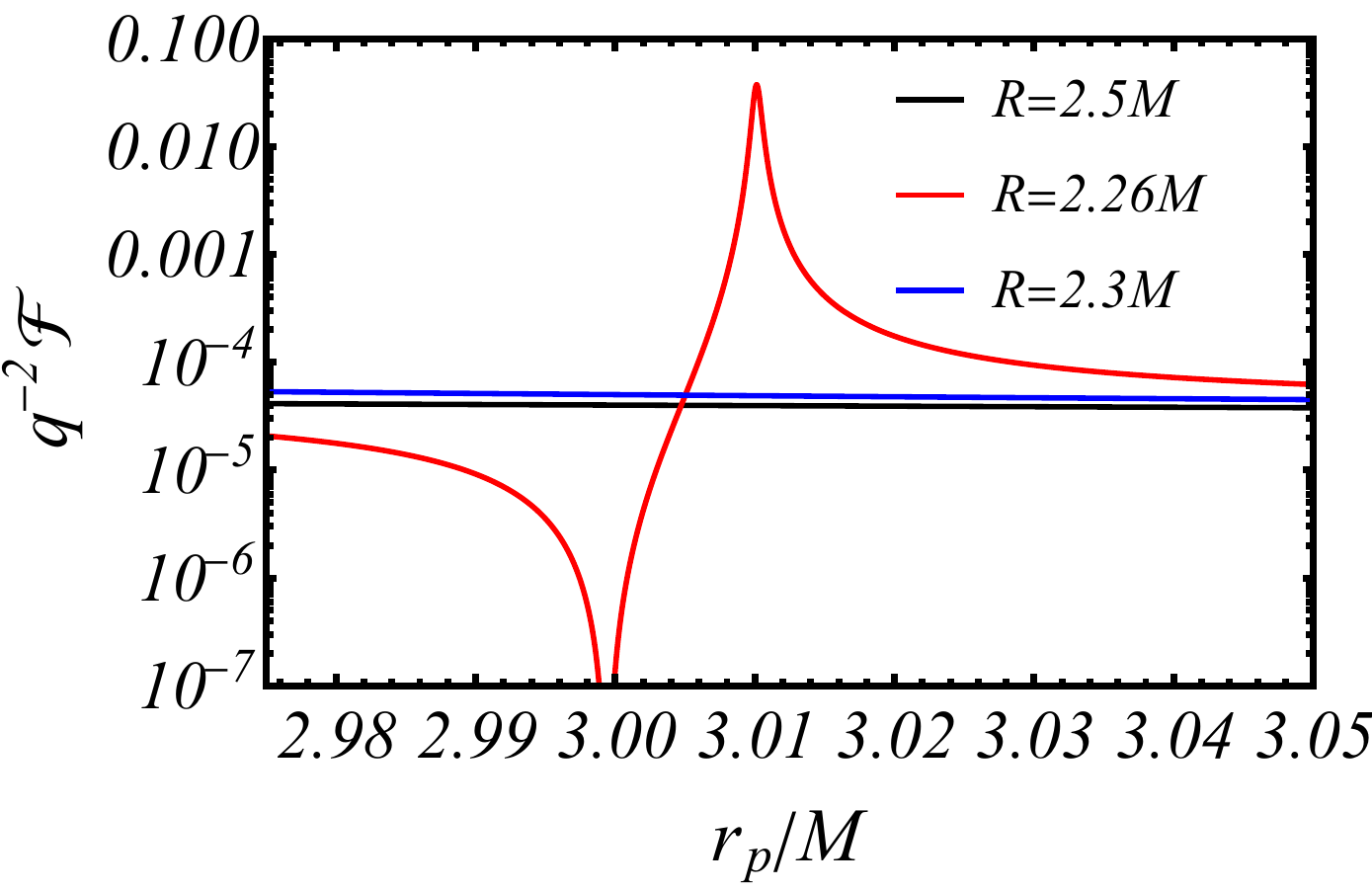}
\end{tabular}
\caption{Energy flux $\mathcal{F}$ emitted in the dipolar mode as a function of the radius of the circular orbit of the point particle. We present results for stars with different compactness. For the very compact configuration with $R=2.26M$, the orbital motion resonantly excites the QNM with frequency $M\omega_\text{QNM}=0.1633 - i 2.470 \times 10^{-5}$ (corresponding to an orbital radius $r_p=3.011M$). Off-resonance, the flux drops several orders of magnitude, which can hamper the inspiral of the particle if only this multipole is considered (this behavior had also been observed in Ref.~\cite{Maggio:2021uge}). The off-resonance values can be used in condition Eq.~\eqref{eq:MassCondition} to estimate how small the mass of the point particle needs to be for it to spend enough time near the QNM frequency so that the resonance fully develops. Note that these are frequency-domain results, for which the particle sits at each $r_p$ without backreaction from the radiation, and they would coincide with those from an adiabatic evolution {\it in the frequency domain}.}
\label{fig:Flux_freq}
\end{figure}
\begin{figure*}[ht!]
\begin{tabular}{c}
\includegraphics[width=0.66\columnwidth]{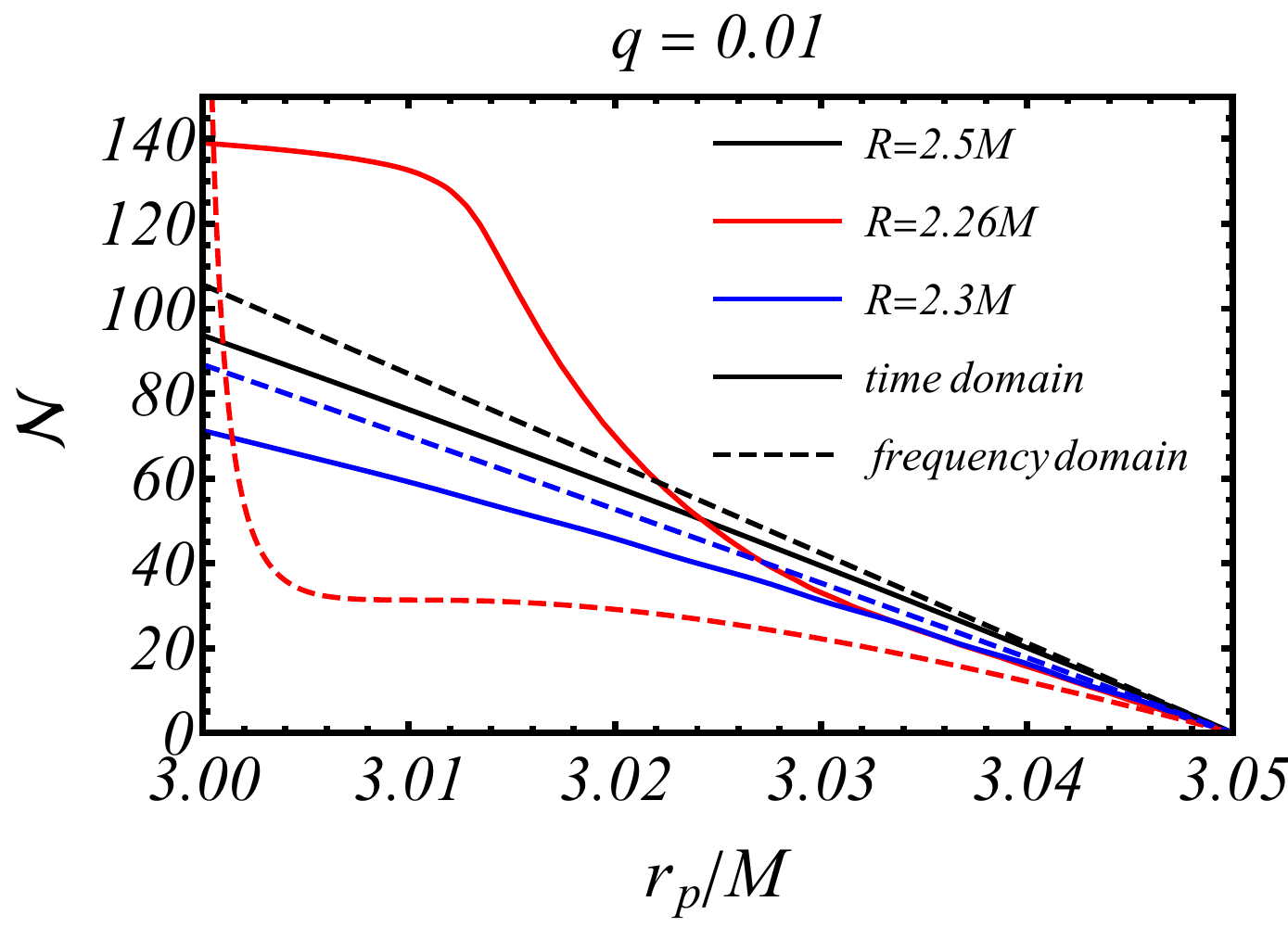}
\includegraphics[width=0.66\columnwidth]{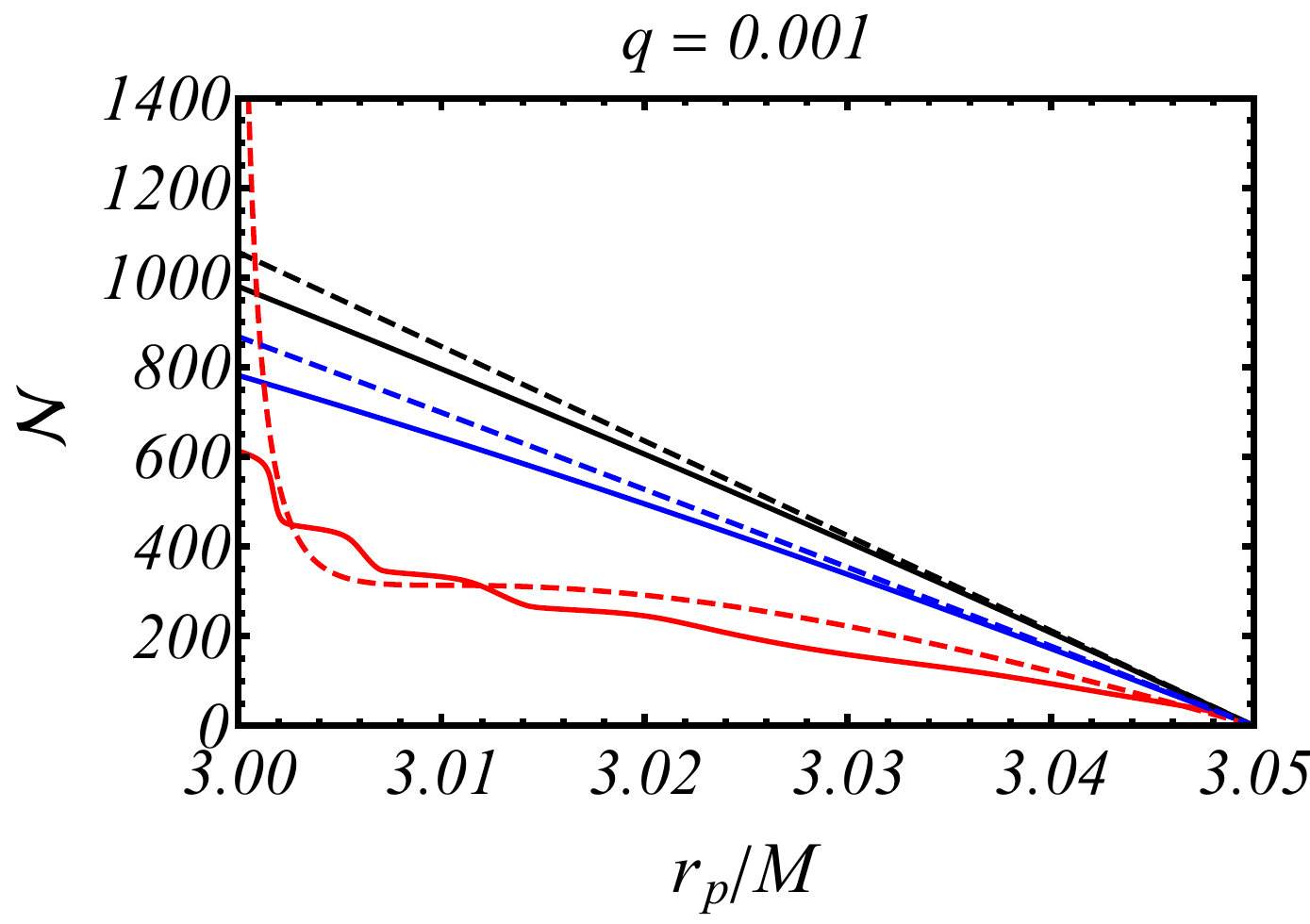}
\includegraphics[width=0.66\columnwidth]{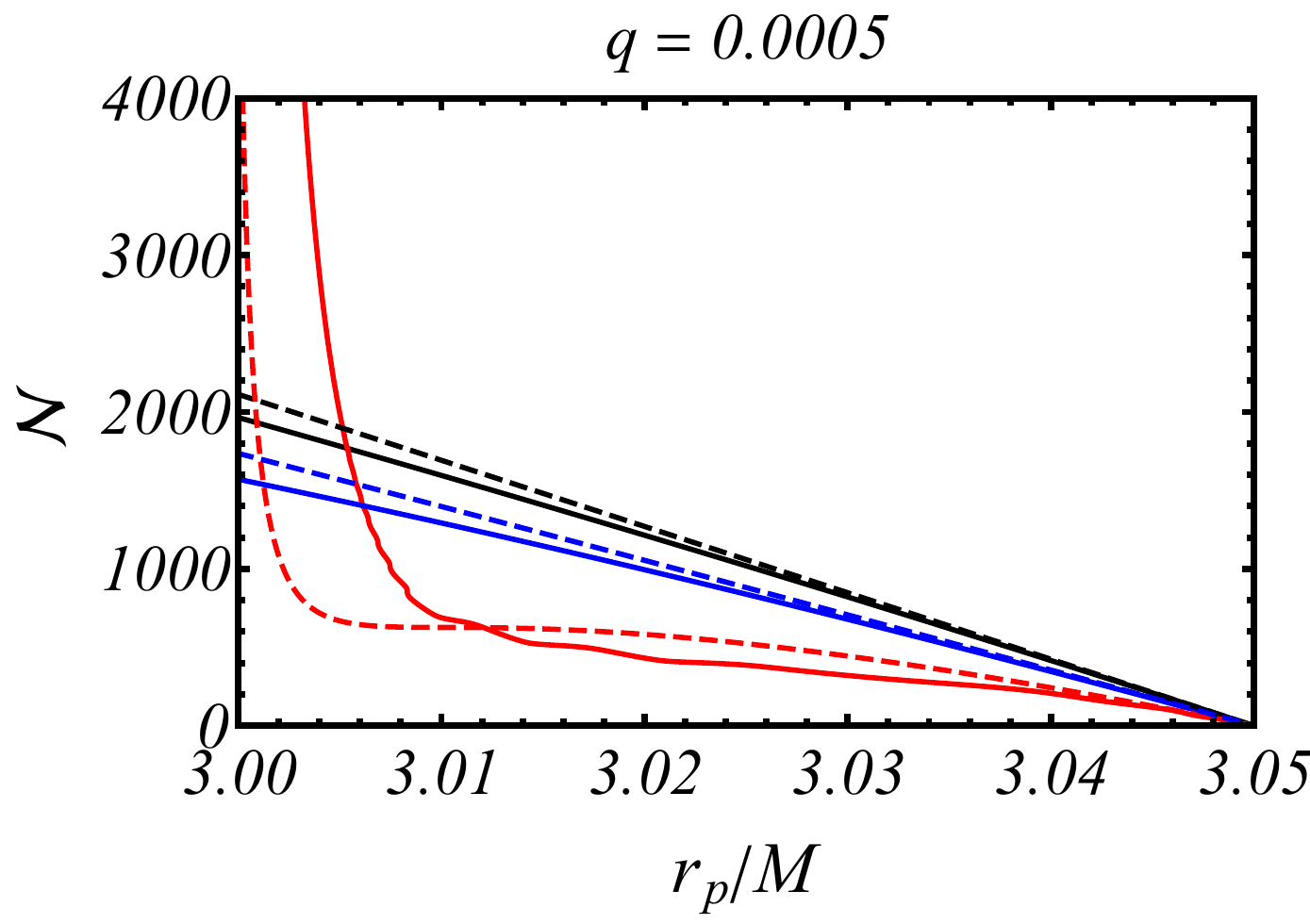}\\
\includegraphics[width=0.66\columnwidth]{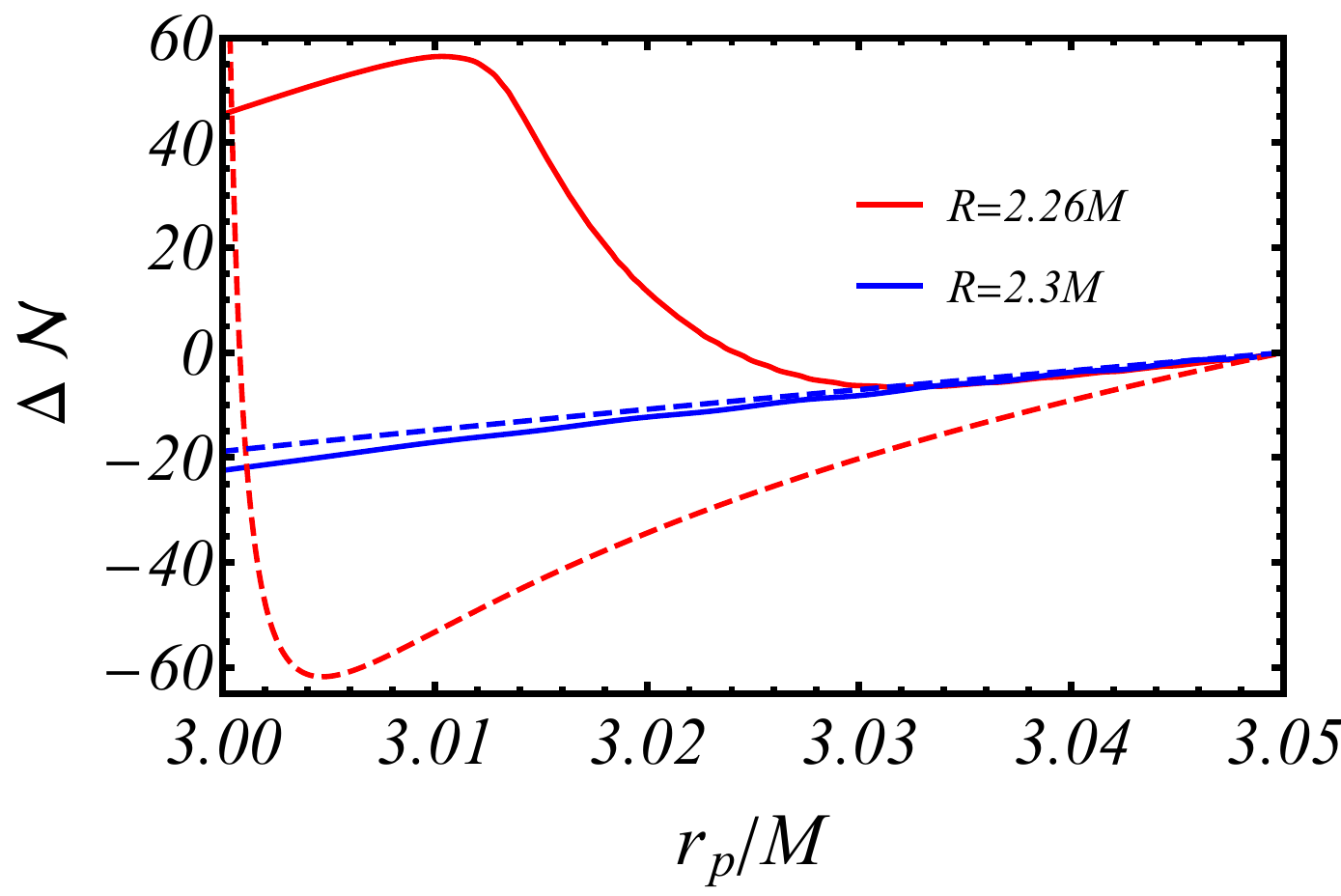}
\includegraphics[width=0.66\columnwidth]{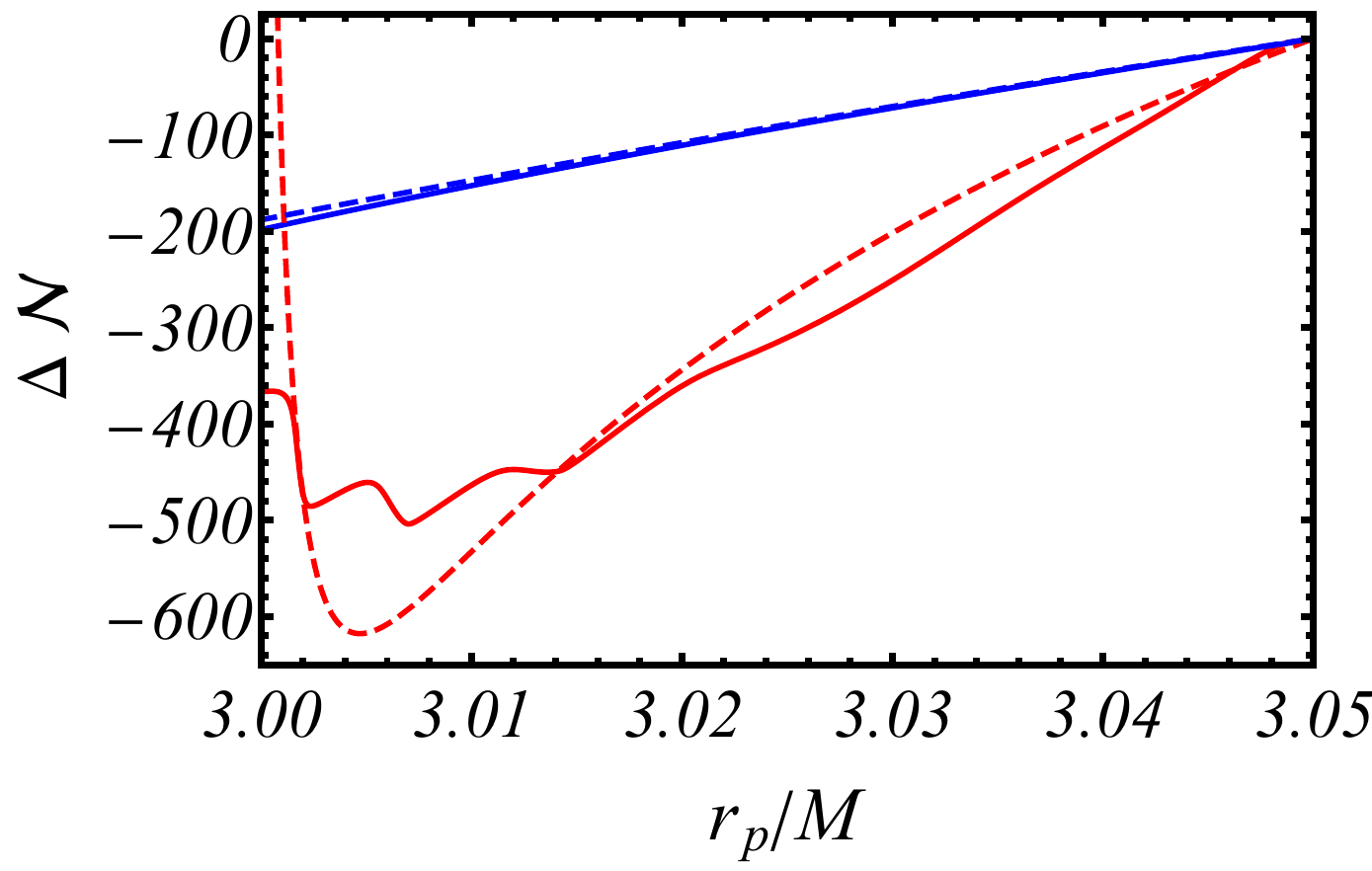}
\includegraphics[width=0.66\columnwidth]{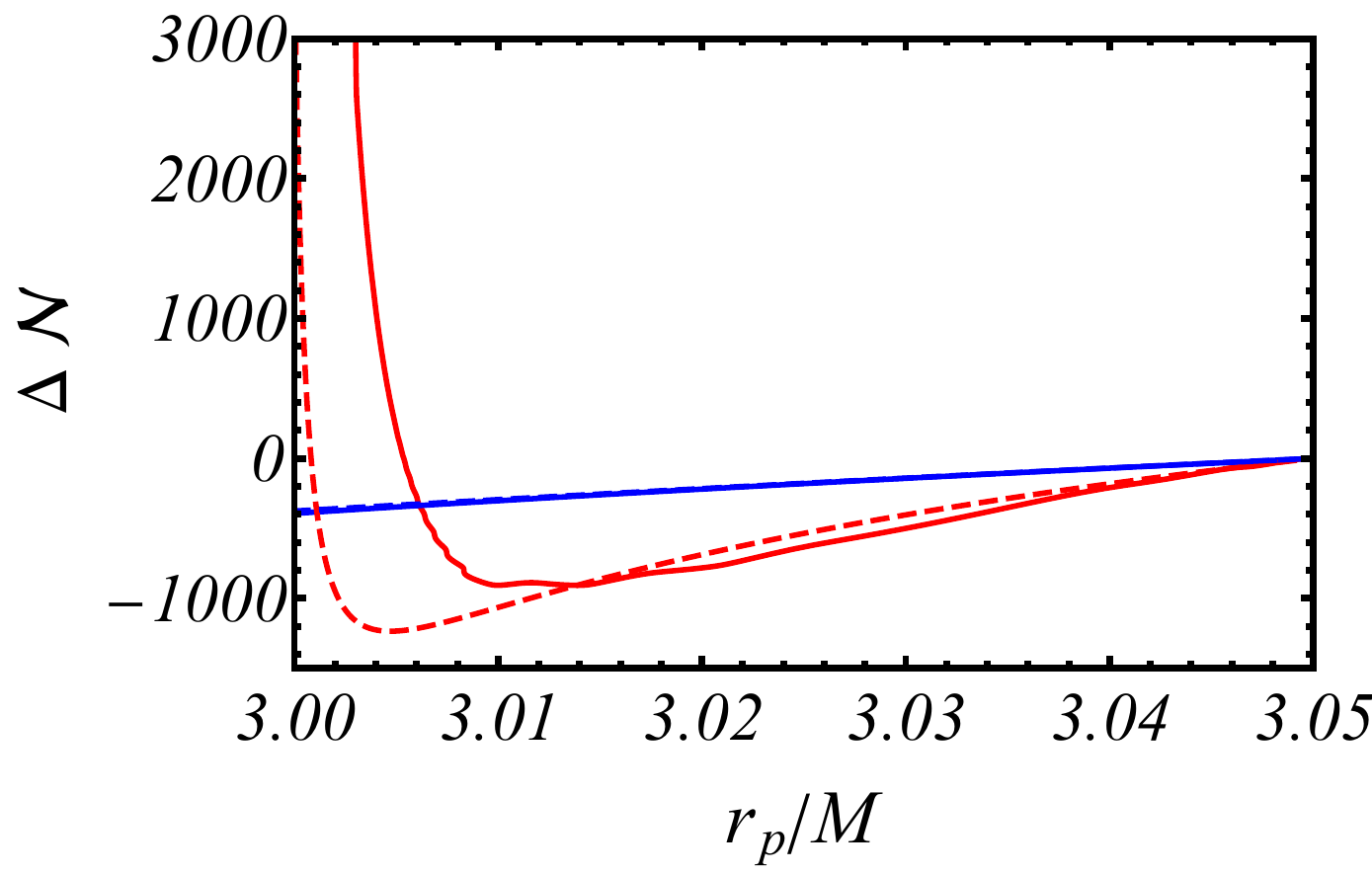}
\end{tabular}
\caption{\textbf{Top row}: Number of cycles $\mathcal{N}$ performed by a point-particle during an inspiral around a constant-density star, as a function of the particle's radius.
The particle starts at $r_p=3.05$ and evolves via radiation reaction in the time- and frequency-domain. Notice how the two results are different close to a resonance, unless the mass of the point particle is very small. \textbf{Bottom row}: Difference in the number of cycles performed by the point particle with respect to those around the less compact star $R=2.5M$, $\Delta \mathcal{N}(R)=\mathcal{N}_R - \mathcal{N}_{2.5M}$. The larger the mass ratio, the more the frequency-domain analysis differs from that in the time-domain, indicating that the stationary assumption -- used by construction in the frequency domain -- is not appropriate to evolve the system. The upper limit mass ratio for (quasi)stationarity of the binary is $q_\text{max}\sim 5\times 10^{-4}$, in agreement with the analytical estimate of Eq.~\eqref{eq:MassCondition}.}
\label{fig:l1_Cyles} 
\end{figure*}

The above results strongly suggest that in order to excite a resonance, the system needs to spend at least a time $\sim 1/\omega_I$ in a frequency band $\delta \omega \sim \omega_I$
around the resonance at $\omega_R$. Rigorous estimates for simple linear differential equations were obtained in Refs.~\cite{Fowler:1921,Nayfeh:book}. We can work out the consequences for GW science: the time $\delta t_\text{cross}$ that the system takes to cross the resonance is~\cite{Cardoso:2019nis}
\be
\delta t_\text{cross} \sim \omega_I/\left(d\Omega/dt \right) \, ,
\ee
with
\be
\frac{d\Omega}{dt} = \left(\frac{d\Omega}{dr}\right)\left(\frac{dr}{dE}\right)\mathcal{F} \,.
\ee
Then, for the resonance to be effectively excited
\beq
&&\omega_I\delta t_\text{cross} \gtrsim 1 \nonumber\\
&\Leftrightarrow& q \lesssim q_\text{max} = \frac{(M \omega_I)^2}{q^{-2} \mathcal{F}}\left( \frac{d\epsilon}{dr}\right)\bigg/\left(M \frac{d\Omega}{dr} \right)  \, . \label{eq:MassCondition} 
\eeq 
In this estimate, orbital quantities on the right hand side are meant to be evaluated at the radius where the resonance is excited, and the flux is to be taken \textit{outside} the resonance, since this is the actual energy flux emitted by the system while the resonance grows. 

Figure~\ref{fig:Flux_freq} shows frequency-domain results for the normalized energy flux emitted in the dipolar mode, as a function of the radius of the point particle. 
For the very-compact star with $R=2.26M$, we can distinguish the narrow resonant peak resulting from the excitation of the proper mode of the star with frequency $M\omega_\text{QNM}=0.1633 - i 2.470 \times 10^{-5}$ (cf. Table~\ref{tab:QNMs}). Frequency-domain calculations {\it assume} stationarity of the field and flux, by construction. For such an assumption to be justified, one needs to full criteria~\eqref{eq:MassCondition}. This implies that, in order to efficiently excite the resonance, the mass of the point particle needs to be 
$m_p \lesssim 10^{-5}-10^{-4}M$.

In order to test the robustness of this estimate, we studied inspiralling trajectories also in the time-domain, therefore not subjected to a stationarity assumption. To simplify both the procedure and the physical interpretation of the results, we only took into account the dipolar mode in the evolution. A useful quantity to evaluate the discrepancy between the two frameworks is the number of orbital cycles $\mathcal{N}$ that the point particle does around the central object
\beq
\mathcal{N}(t) = \frac{\varphi(t)}{2\pi} \,,
\eeq
as it sweeps some fixed frequency band, parametrized by its initial and final orbital radius.

Our findings are summarized in Fig.~\ref{fig:l1_Cyles}. We show both the absolute value of the number of cycles for different mass ratios, as well as the relative difference  $\Delta \mathcal{N}$ with respect to the number of cycles performed around the less compact star with radius $R=2.5M$. For all cases, we start the motion at radius $r_p=3.05M$, which corresponds to an orbital frequency of 
$\Omega=0.1606  \approx \omega_\text{R} - 110\,\omega_I$. In the time-domain, we wait a time $t_\text{start}\sim 2000M$ before starting the inspiral to let the initial burst of energy be washed away, and for the system to reach a (quasi)stationary state.

For stars where no resonance is excited ($R=2.3M$ and $R=2.5M$), the frequency-domain results show a relative difference of $\sim 5 \%$ to the time-domain ones at the end of our monitoring of the binary. This discrepancy is independent of the mass ratio and can be attributed to small numerical differences accumulated over many orbits. Moreover, for the $R=2.3M$ configuration we would need to wait much longer to start the time-domain inspiral from a complete stationary state. As expected from Fig.~\ref{fig:Flux_freq}, the inspiral for the more compact geometry performs less cycles, since the flux is larger so the binary is losing energy faster. Also, the  number of cycles is scaling with the inverse of the mass ratio, which is consistent with the typical inspiral timescale  $t_\text{inspiral} \propto M^2/m_p$.

For the very-compact configuration $R=2.26M$, where a resonance can be excited, the differences are more sounding. The frequency-domain analysis again presents a scaling with the inverse of the mass ratio. Clearly, this does not hold for the time-domain analysis. For the smallest mass ratio presented, $q= 0.0005$, the shape of the time-domain curves are similar to the frequency-domain ones, which indicates the binary is evolving in a quasi-stationary state. The resonance is properly excited and the particle falls in the flux ``well'' that comes after the resonance peak (cf. Fig.~\ref{fig:Flux_freq}), which stalls the inspiral. Consequently, the number of cycles increases at a fixed radius as if the particle was ``frozen'' there. This stalling would be less significant if we were taking into account higher multipoles. 

However, as the mass of the point particle increases so do the differences between the time and frequency-domain analysis. For the intermediate mass ratio of $q=0.001M$, the inspirals have similar behaviors initially, but the resonance is less pronounced in the time-domain, since the particle does not spend enough time there for it to totally grow. Similarly, the particle also does not get stuck at the ``zero'' of the frequency-domain flux as before. The small wiggles observed can be attributed to the oscillations in the flux as the the system relaxes to stationarity, as discussed in the previous section. Since in this configuration the relaxation timescale is comparable to the inspiral timescale, the wiggles become noticeable. Finally, for the highest mass ratio considered, $q=0.01M$, the particle quickly crosses the resonance, and this never develops. As a consequence, the time-domain results disagree completely with the frequency-domain. We conclude that for the system studied, the adiabatic approximation employed in the frequency-domain is only appropriate when the mass ratio is smaller than $q_\text{max} \sim 5 \times 10^{-4}$. This upper bound agrees with the estimate given in Eq.~\eqref{eq:MassCondition}.

\section{Discussion}
%
\begin{figure}[t]
\begin{tabular}{c}
\includegraphics[width=0.95\columnwidth]{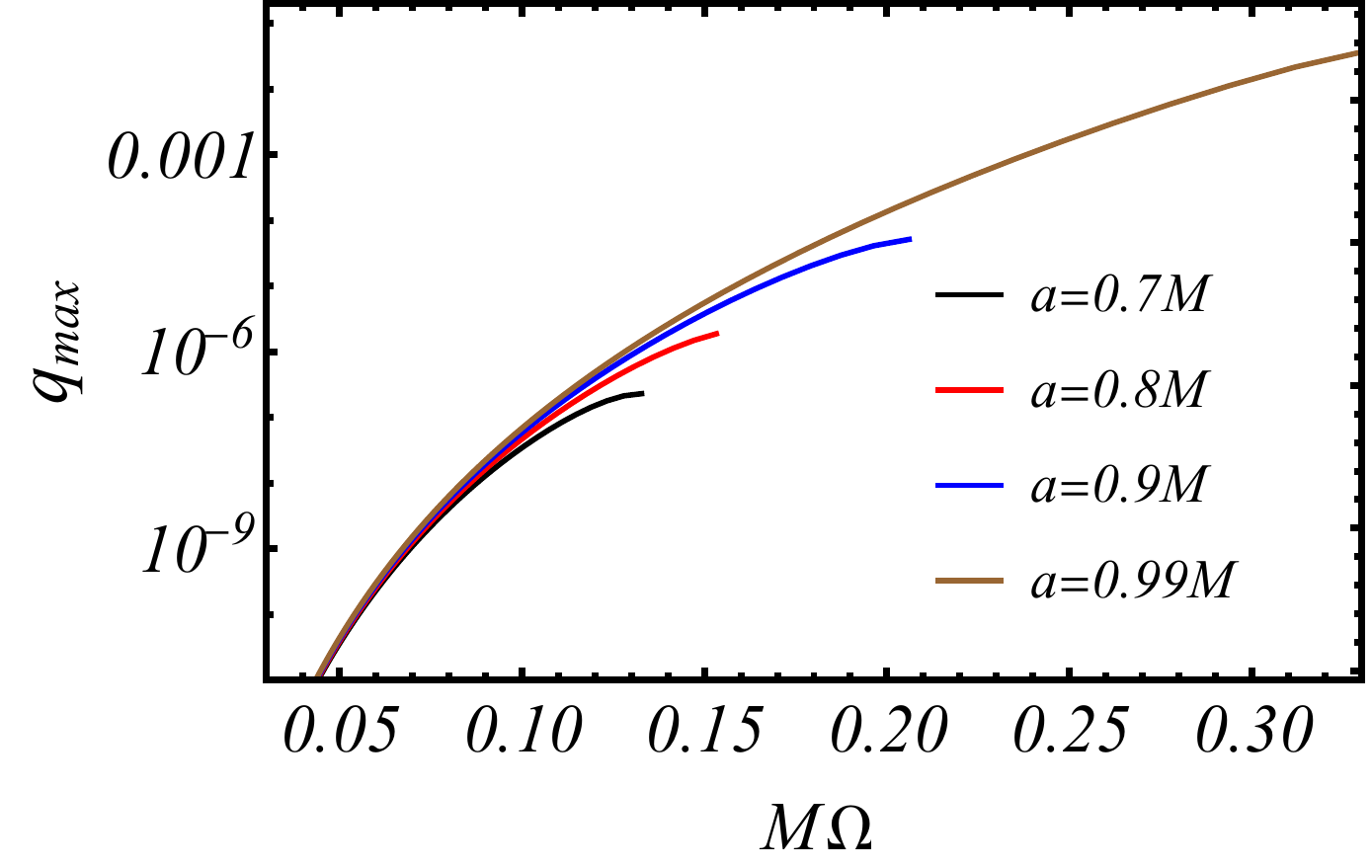}
\end{tabular}
\caption{Largest mass ratio predicted by the estimate~\eqref{eq:MassCondition} that would allow the resonant excitation, by GWs, of a QNM of an exotic compact object with frequency $\omega_R = 2\Omega$. We considered that $\omega_I \sim \omega_R^{2l+3}$ \cite{Cardoso:2019nis,Maggio:2021uge} and only took into account the quadrupolar mode in the energy flux, making this estimate conservative, since $\mathcal{F}$ will increase if higher multipoles are considered. We show orbital frequencies $\Omega$ corresponding to radius of the particle from $r_p=10M$ up to  almost the ISCO for each spin. For mass ratios larger than these limits, the particle crosses the resonance too quickly for it to effectively grow.}
\label{fig:qmax}
\end{figure}
We have shown is that spacetimes with photonspheres behave as cavity resonators in electromagnetism~\cite{Milton:2006ia,PhysRevLett.110.237401}. They have large ``build-up'' times, dictated by the transmission amplitudes at the photonsphere, and are prone to resonances.
A proper accounting of the evolution of such systems demands that one takes into account the energy piling up within the photonsphere. For systems evolving rapidly under radiation reaction, 
such horizonless objects are effectively absorbers and should mimic BHs well indeed. A proper modeling of this process, and the full evolution of an extreme mass ratio system is an open problem.

Such systems also have resonances, which can be probed by orbiting bodies. However, we argued that, for the resonance to fully develop in a binary, the latter has to evolve slower than the time needed for the resonance to grow. Such a condition imposes stringent constraints on possible observational tests.

In this work,  we focused on a simple toy-model of a scalar field around a constant-density star, and considered an artificial motion for the point-particle that is not dictated by its equations of motion. Nonetheless, all our analytical estimates for the relaxation timescale \eqref{scattering_prediction}, resonance growth time, and upper limit on the mass ratio that excites a resonance in an inspiralling binary \eqref{eq:MassCondition} are model indepedent and agree with the numerical results for the particular system that we studied. Therefore, our conclusions should be applicable to any astrophysical system. 

Refs.~\cite{Maggio:2021uge, Sago:2021iku} studied extreme-mass-ratio inspirals ($q<10^{-4}$) around spinning horizonless compact objects (see also \cite{Cardoso:2019nis,Fransen:2020prl} for non-spinning analysis and \cite{Fang:2021iyf, Cardoso:2021vjq} for hierarchical triple systems). As in our toy-model, the low-frequency QNMs of the spinning exotic compact object can be resonantly excited during the inspiral, which leads to non-negligible effects in the waveform that need to be considered for the detection and parameter estimation of these sources. However, they work in the frequency-domain and ignore the growth time of the resonance, implicitly assuming stationarity at all instants. As concluded, such approximation is only correct when the mass ratio of the system obeys the condition of Eq.~\eqref{eq:MassCondition}.  

In Fig.~\ref{fig:qmax}, we apply this estimate to the type of systems studied in Refs.~\cite{Maggio:2021uge, Sago:2021iku}. Typically, the resonance width for these exotic compact objects is $\delta \omega \sim \omega_I \sim \omega_R^{2l+3}$ \cite{Cardoso:2019nis, Maggio:2021uge}. In a binary system, the frequency of the emitted GWs is determined by the orbital frequency, and for circular orbits corresponds to $\omega_\text{GW} = 2\Omega$. Then, for every orbital frequency $\Omega$ (or radial location of the particle), we can compute how light the point particle needs to be in order to resonantly excite an ECO with a QNM mode of frequency $\omega_R = 2\Omega$, $\omega_I \sim (2\Omega)^{2l+3}$. For the off-resonance flux $\mathcal{F}$, we used the same values as in Kerr, since the relative difference with respect to an horizonless ultracompact object should be small (though non-negligible when accumutaled over many orbits). We only took into consideration the quadrupolar mode $l=2$, and higher multipoles will typically increase $\mathcal{F}$, therefore placing even more stringents limits on the mass ratio. We conclude that for the reference value of $q=3 \times 10^{-5}$ used in most results presented in Ref.~\cite{Maggio:2021uge}, the particle would only be able to excite resonances in ECOs with spins $a>0.9M$, and on a limited region of the parameter space where it is very close to the central object. 

Our conclusions can also be applied to massive scalar theories \cite{Arvanitaki:2009fg, RevModPhys.84.671,deRham:2014zqa}, where matter orbiting a Kerr BH can resonantly excite superradiant modes, which might lead to so called \textit{floating orbits} \cite{Press:1972zz, Brito:2015oca}. In these orbits, the energy absorbed by the horizon is positive and counterbalances the loss of energy to infinity \cite{Cardoso:2011xi,Yunes:2011aa}. As a consequence, the inspiral freezes and the radiated energy is solely provided by the rotational energy of the BH. These resonances occur for $\omega^2_\text{res} = \mu_s^2 - \mu_s^2 \left(M \mu_s /(l+1+n) \right)^2$ , where $\mu=m_s/\hbar$ is the reduced mass of the scalar field, and have typical widths of $\delta \omega \sim \omega_I \propto \mu_s^{4l+5}$ \cite{Cardoso:2011xi, Detweiler:1980uk}. These are even more narrow than the QNMs of ECOs just discussed. Generically, the off-resonance energy flux is dominated by GWs, which means that for the same orbital frequency one would need even smaller mass ratios than the ones in Fig.~\ref{fig:qmax} in order to properly excite superradiant resonances of massive scalars. 

One might question if additional dissipation mechanisms could undermine the fueling-up of the cavity and excitation of resonances. GWs are known to interact very weakly with matter, with effects only being relevant at the Hubble timescale \cite{1971ApJ...165..165E, 1985ApJ...292..330P, Kocsis:2008aa, Loeb:2020lwa}. Hence, any additional channel of dissipation should be subdominant with respect to the emission of waves to infinity and the trapping of energy by the central object on the timescales of interest for these systems. We cannot rule out, however, that extremely stiff equations of state giving rise to large viscosities and large sound speeds strongly suppress resonances in compact objects. Even in such case, our results still apply to other systems, including resonances of massive boson fields around spinning BHs.

The conclusions of this work have obvious implications to GW astronomy, since a large class of binaries might not be able to probe mechanisms indicative of new physics as effectively as suggested by previous studies employing a frequency-domain analysis, which have largely dominated GW modelling in EMRIs. Being the latter one of the most important scientific targets for the upcoming space-based LISA mission \cite{Barack:2018yly, Barausse:2020rsu}, in particular to test GR and hunt for new fundamental physics, we hope our work draws the attention of the community to the necessity of better understanding GW emission in less conventional systems that are not typical binaries in GR.

\noindent{\bf{\em Acknowledgments.}}
%
We are grateful to Rodrigo Vicente, for fruitful discussions and warm hospitality while this work was finalized. 
We are also indebted to Adrian del Rio, Elisa Maggio,  Maarten van de Meent, Paolo Pani and Takahiro Tanaka for useful feedback and comments on a first version of this manuscript.
This work makes use of the Black Hole Perturbation Toolkit. 
V. C. is a Villum Investigator supported by VILLUM FONDEN (grant no. 37766) and a DNRF Chair supported by the Danish Research Foundation.
F. D. acknowledges financial support provided by FCT/Portugal through grant No. SFRH/BD/143657/2019. 
This project has received funding from the European Union's Horizon 2020 research and innovation programme under the Marie Sklodowska-Curie grant agreement No 101007855.
We thank FCT for financial support through Project~No.~UIDB/00099/2020.
We acknowledge financial support provided by FCT/Portugal through grants PTDC/MAT-APL/30043/2017 and PTDC/FIS-AST/7002/2020.
%

\appendix

\section{Unstable circular geodesics}\label{sec:Unstable}
\begin{figure*}[ht!]{}
\begin{tabular}{c}
\includegraphics[width=0.66\columnwidth]{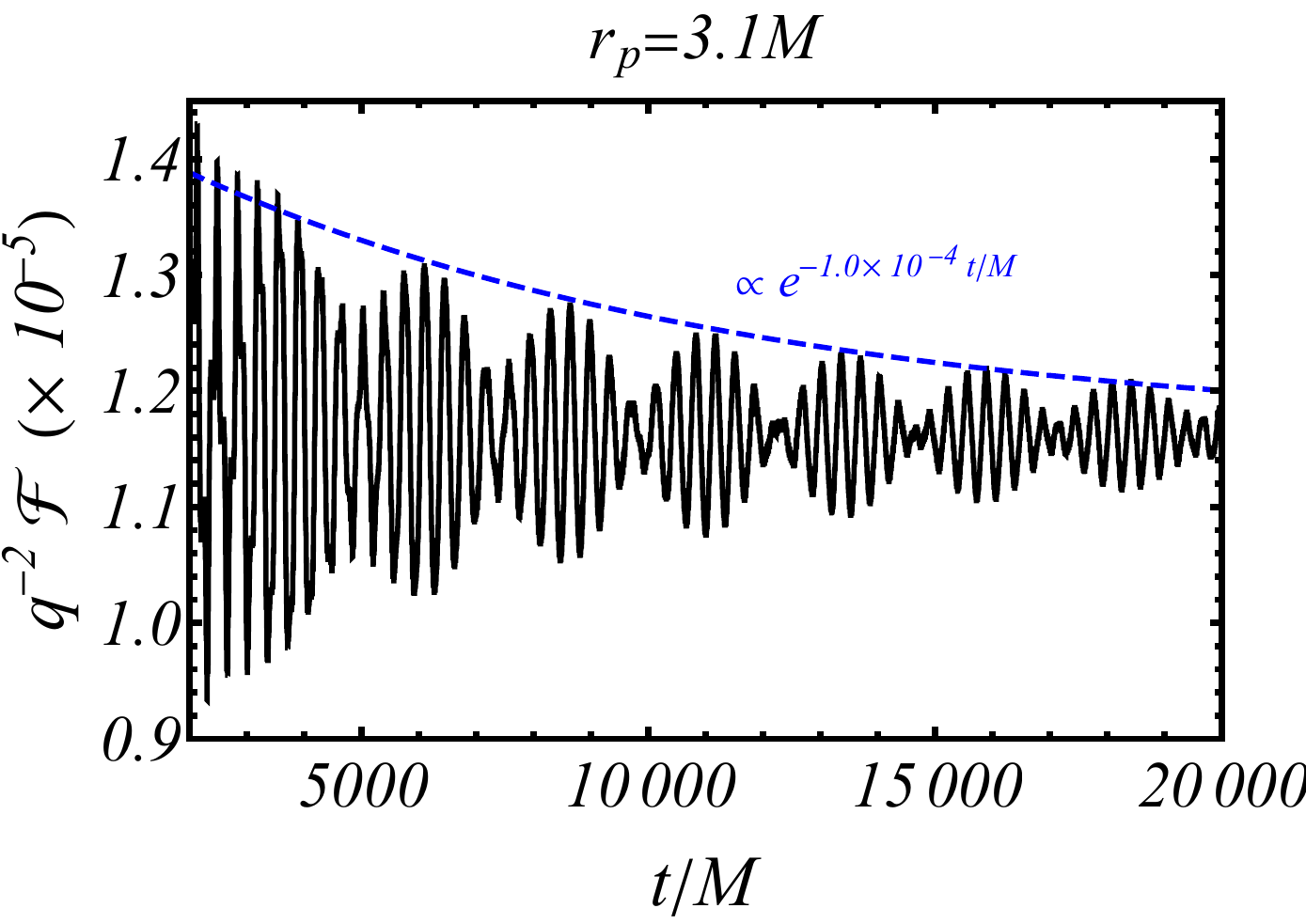}
\includegraphics[width=0.66\columnwidth]{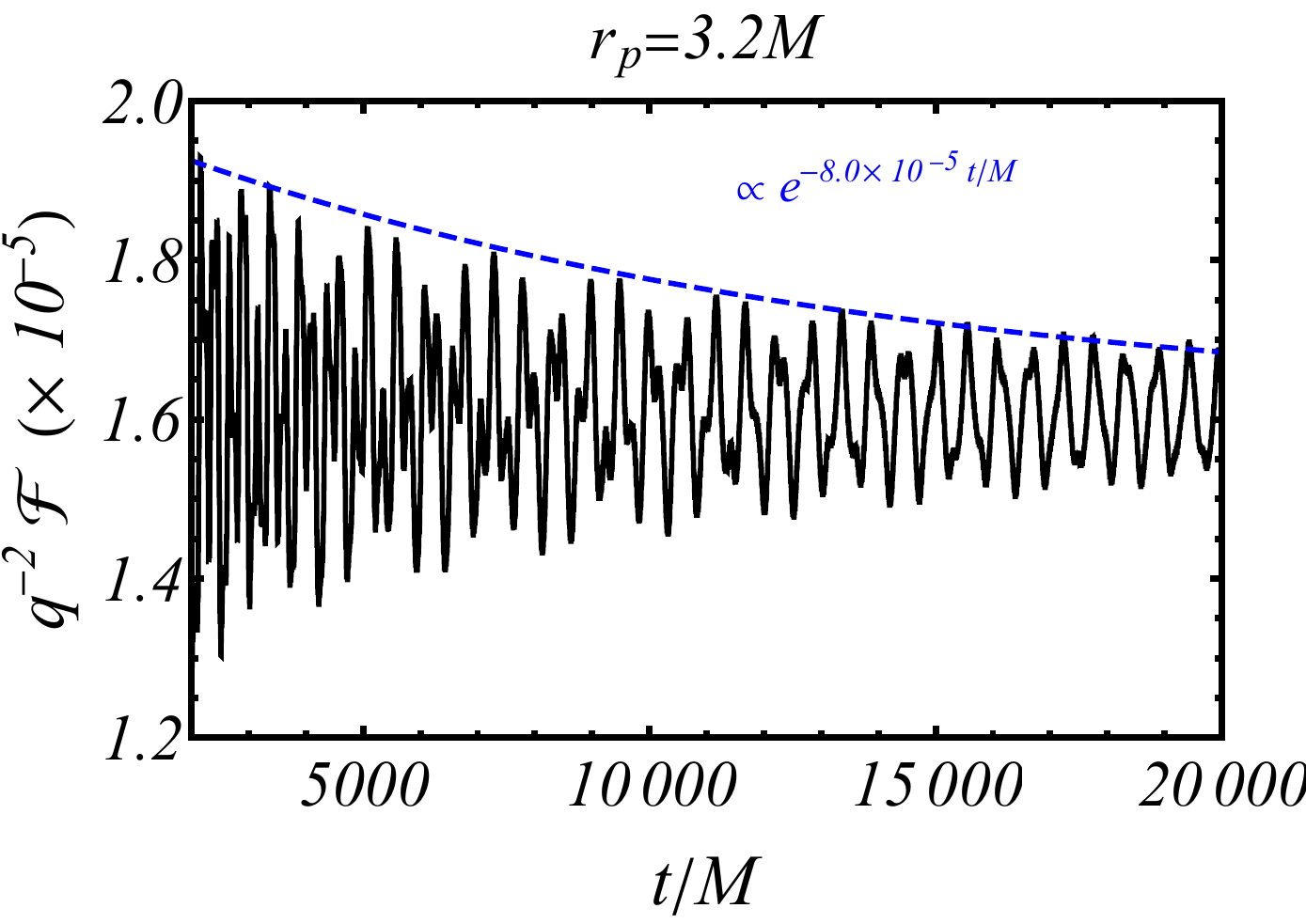} \\
\includegraphics[width=0.66\columnwidth]{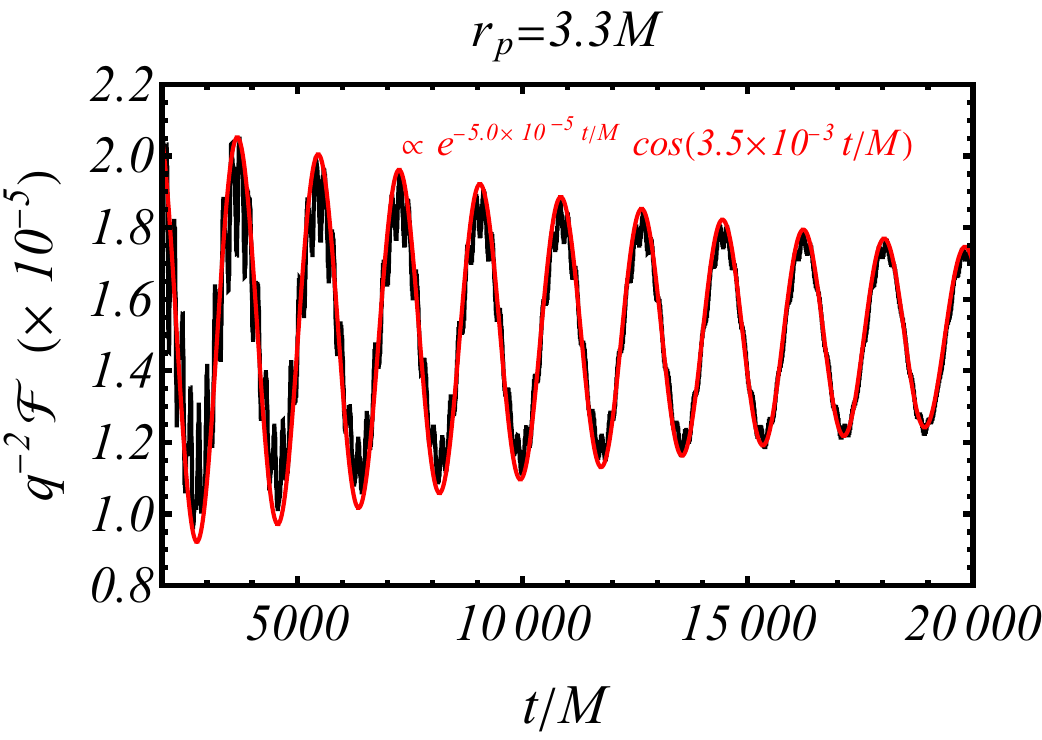} 
\includegraphics[width=0.66\columnwidth]{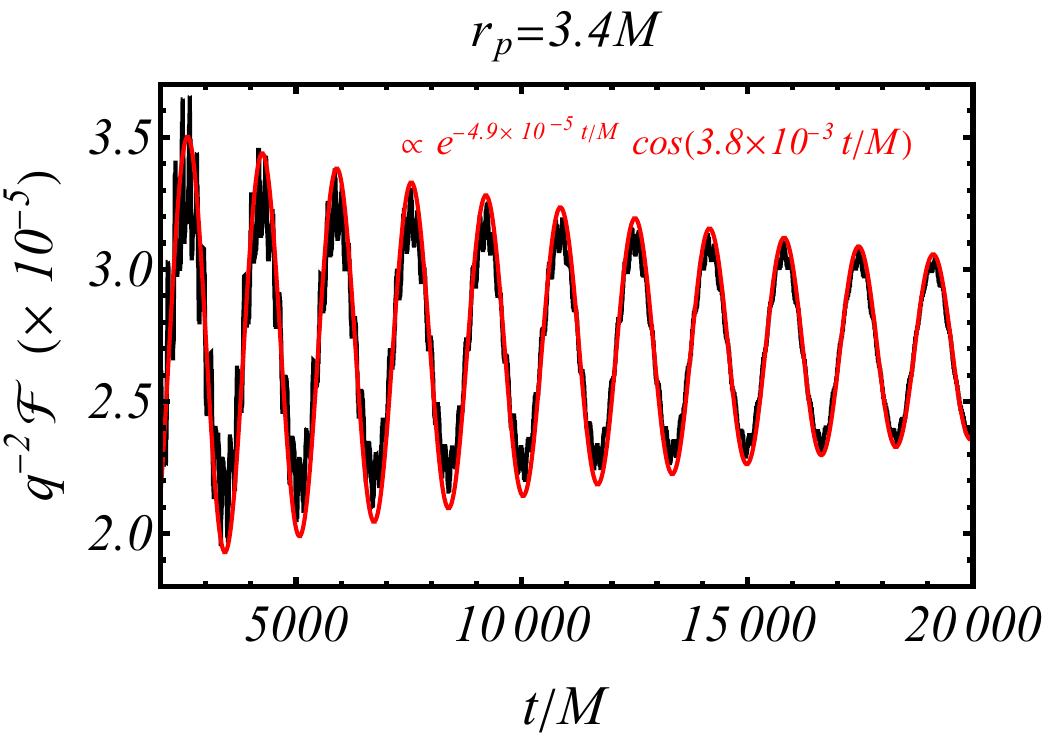} 
\includegraphics[width=0.66\columnwidth]{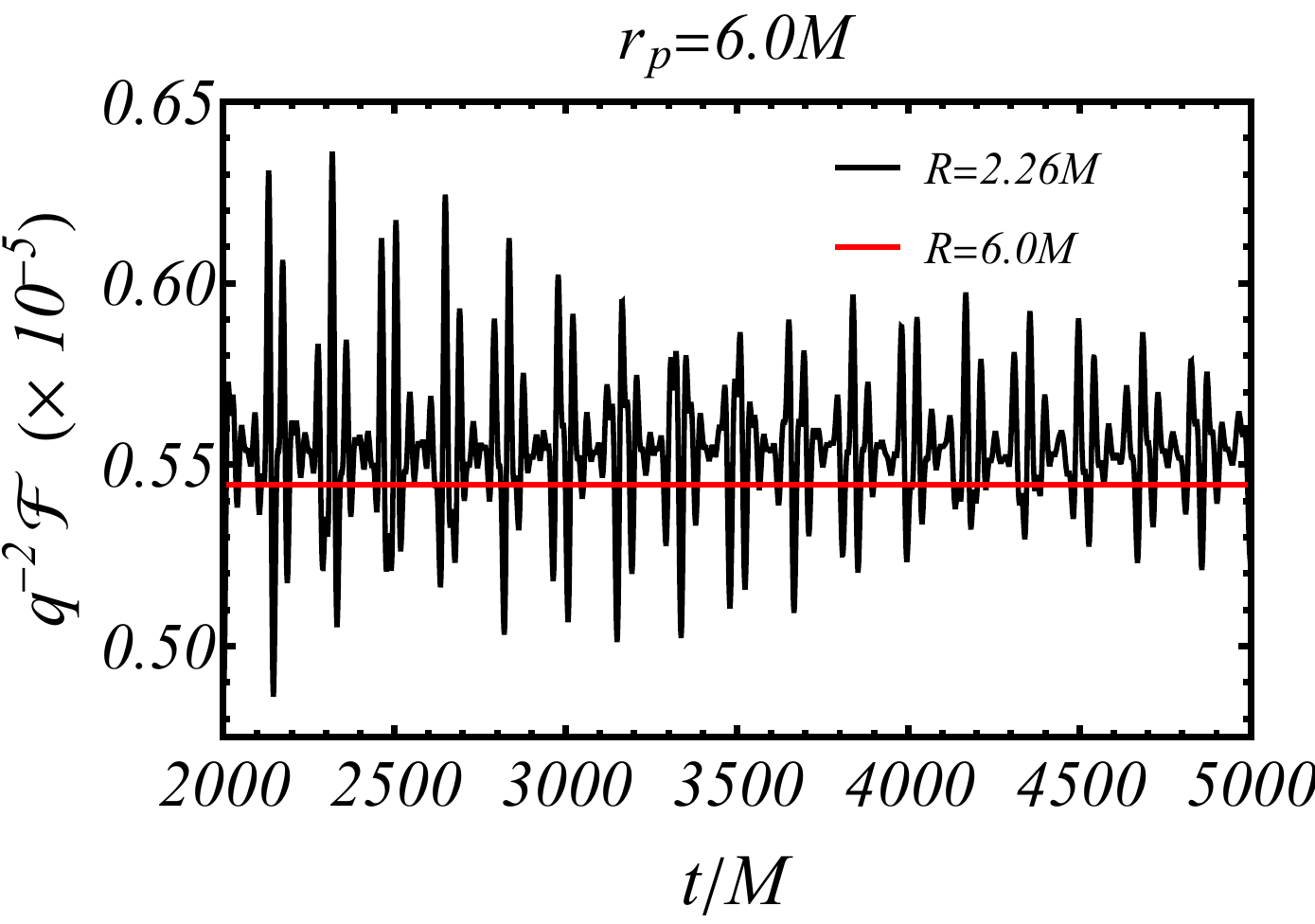}\\
\end{tabular}
\caption{Same analysis as in Fig.~\ref{fig:l1} but considering circular geodesics around Schwarzschild, i.e. $a=0M$ in Eq.~\eqref{eq:AngularFreq}-\eqref{eq:Lz}. We do not show an analogous plot for the $M\omega=0.192$ since the particle would have to be put at the light-ring ($r_p=3.0M$) and we are only considering timelike motion.
}
\label{fig:l1_a0} 
\end{figure*}

In this Appendix, we present results similar to the ones discussed in Section \ref{sec:buildup_numerical} but considering the point particle is on circular geodesics, i.e. with $a=0M$ in Eqs.~\eqref{eq:AngularFreq}-\eqref{eq:Lz}. Our results are summarized in Fig.~\ref{fig:l1_a0}. We put the particle at a radius that yields the same frequency as those presented in Fig.~\ref{fig:l1}. Apart from a change in the absolute values for the flux, the relaxation timescales are in complete agreement with the ones obtained before, indicating the artificial motion we considered there is irrelevant to the excitation of the constant-density star's QNMs. As previously mentioned, to excite the fastest growing QNM of the star (see Table~\ref{tab:QNMs}) with geodesic motion we would have to put the particle at unstable circular orbits, i.e. $r_p < r_\text{ISCO}=6.0M$ for $a=0M$, which would prevent us from studying inspiralling trajectories. 

\bibliographystyle{h-physrev4}
\bibliography{references} 
\end{document}